\documentclass[prl,twocolumn,showpacs,superscriptaddress]{revtex4-1}
\usepackage{amsfonts}
\usepackage{amsmath}
\usepackage{amsthm}
\usepackage{braket}
\usepackage{graphicx}
\usepackage{hyperref}
\usepackage{color}
\usepackage[utf8]{inputenc}
\usepackage{dsfont}

\newcommand{\simplesection}[1]{\emph{#1} ---}
\newcommand{\peps}{\ensuremath{A}}

\begin{document}

\title{Characterizing Topological Order with Matrix Product Operators}

\author{Mehmet Burak \surname{\c{S}ahino\u{g}lu}}
\affiliation{Vienna Center for Quantum Technology, University of Vienna, Boltzmanngasse
5, 1090 Vienna, Austria}
\author{Dominic \surname{Williamson}}
\affiliation{Vienna Center for Quantum Technology, University of Vienna, Boltzmanngasse
5, 1090 Vienna, Austria}
\author{Nick \surname{Bultinck}}
\affiliation{Department of Physics and Astronomy, University of Ghent, Krijgslaan 281
S9, B-9000 Ghent, Belgium}
\author{Michael \surname{Mari\"en}}
\affiliation{Department of Physics and Astronomy, University of Ghent, Krijgslaan 281
S9, B-9000 Ghent, Belgium}
\author{Jutho \surname{Haegeman}}
\affiliation{Department of Physics and Astronomy, University of Ghent, Krijgslaan 281
S9, B-9000 Ghent, Belgium}
\author{Norbert \surname{Schuch}}
\affiliation{JARA Institute for Quantum Information, RWTH Aachen University, D-52056 Aachen, Germany}
\author{Frank \surname{Verstraete}}
\affiliation{Vienna Center for Quantum Technology, University of Vienna, Boltzmanngasse
5, 1090 Vienna, Austria}
\affiliation{Department of Physics and Astronomy, University of Ghent, Krijgslaan 281
S9, B-9000 Ghent, Belgium}

\begin{abstract}
One of the most striking features of quantum phases that exhibit topological order is the presence of long 
range entanglement that cannot be detected by any local order parameter. The formalism of projected entangled-pair 
states is a natural framework for the parameterization of the corresponding ground state wavefunctions, in which 
the full wavefunction is encoded in terms of local tensors. Topological order is reflected in the symmetries of 
these tensors, and we give a characterization of those symmetries in terms of matrix product operators acting 
on the virtual level. This leads to a set of algebraic rules characterizing states with topological quantum order. 
The corresponding matrix product operators fully encode all topological features of the theory, and provide a 
systematic way of constructing topological states. We generalize the conditions of $\mathsf{G}$ and twisted injectivity 
to the matrix product operator case, and provide a complete picture of the ground state manifold 
on the torus. As an example, we show how all string-net models of Levin and Wen fit within this formalism, and 
in doing so provide a particularly intuitive interpretation of the pentagon equation for F-symbols as the 
pulling of certain matrix product operators through the string-net tensor network. Our approach paves 
the way to finding novel topological phases beyond string-nets, and elucidates the description of topological phases 
in terms of entanglement Hamiltonians and edge theories.
\end{abstract}

\maketitle

Classifying phases of matter is one of the most important problems in condensed matter physics. Landau's 
theory of symmetry breaking \cite{Landau75} has been extremely successful in characterizing phases in terms 
of local order parameters, but it has been known since the work of Wegner \cite{Wegner71} that topological 
theories do not necessarily exhibit such a local order parameter, and hence that  different topological phases 
cannot be distinguished locally. One of the main reasons to call such phases topological is the fact that the 
ground state degeneracy depends on the topology of the surface on which the system is defined \cite{Wen90}. 
Since the realization that quantum Hall systems exhibit topological quantum order \cite{WenNiu90}, significant 
effort has been put into classifying all topological phases \cite{WenZee92, Kitaev06, Schnyder08, FidkowskiKitaev11, XieGuWen11, SchuchGarciaCirac11}. 
A very large class of models exhibiting topological order was constructed by Levin and Wen \cite{LevinWen05}, 
which is conjectured to provide a complete characterization of non-chiral topological theories in two dimensions.

A recent development at the interface of quantum information and condensed matter theory is the 
growing use of projected entangled-pair states (PEPS), and more general tensor network states \cite{Fannes92, Klumper93, VerstraeteMurgCirac08}. 
To construct a PEPS one associates a tensor, representing a map from some
virtual vector space to the local physical Hilbert space, to each site of
a lattice and performs tensor contractions on the virtual space according
to the graph of the lattice. The resulting quantum state can then be used
as an ansatz for the ground state of a local Hamiltonian on that lattice \cite{VerstraeteCirac06, Hastings07, GarciaVerstraeteWolfCirac08}.
There are two immediate and very important properties of PEPS. Firstly, 
for every PEPS there exists a local, positive-semidefinite, frustration free operator called the parent 
Hamiltonian whose kernel contains the PEPS. Secondly, the entanglement entropy of a 
region $R$ is upper bounded by $|\partial R| \log D$  rather than the volume, where $D$ is the virtual 
dimension and $|\partial R|$ is the number of virtual bonds crossing the boundary of the region.  
Hence PEPS are the ground states of local Hamiltonians and obey an area law (provided the bond dimension 
is upper bounded by a fixed constant $D$ as the system size increases).

In this Letter, we propose a general framework for the exact description of topologically ordered ground 
states using tensor networks. To achieve this we generalize the concept of $\mathsf{G}$ injectivity \cite{SchuchCiracGarcia10}, 
where the relevant subspace of the virtual space is invariant under a tensor product action of some 
symmetry group $\mathsf{G}$, and its extension to twisted group actions~\cite{Buerschaper14}. The 
generalized notion of injectivity presented in this Letter provides a natural extension of these concepts 
that applies even when no group symmetry is involved. Furthermore, it allows for the consistent characterization 
of the invariant subspace on the virtual level across arbitrary lattice bipartitions  in terms of local tensors 
that form a projection matrix product operator (MPO).
Extending beyond the group case is significant as it is necessary for the description of more general 
topological orders including the string-net models.

We first define MPO injectivity, proceed by formulating a set of algebraic conditions that have 
to be satisfied by valid MPOs, and then show how the ground state degeneracy and topological order is 
determined by those MPOs. We go on to illustrate that all ground states of the string-net models satisfy 
the proposed algebraic conditions, and that the key \emph{pulling through} condition for these models 
is implied by the pentagon equation for the $F$-symbols. We conclude by providing an outlook towards 
possible extensions of the framework to fermionic models and higher dimensional theories, and a 
discussion of the potential relevance of our formalism to the development of more efficient PEPS contraction schemes.

\simplesection{MPO injectivity}
Consider a tensor $\peps$ for building a PEPS on a lattice with coordination number three, which can be 
understood as a linear map from the virtual space to the physical space, $\peps: \mathbb{C}^D  \otimes \mathbb{C}^D \otimes \mathbb{C}^D \rightarrow \mathbb{C}^d$. 
By blocking several tensors, we can define a mapping from the virtual boundary space 
$(\mathbb{C}^D)^{\otimes |\partial V|}$ to the physical bulk $(\mathbb{C}^{d})^{\otimes |V|}$ for arbitrary 
regions $V$ of the lattice.
There always exists some maximal subspace $S_V$ of the virtual space $(\mathbb{C}^D)^{\otimes |\partial V|}$ 
on which this mapping is injective. Denoting the projection onto $S_V$ by $\openone_{S_V}$, the key 
condition of MPO injectivity is that $\openone_{S_V}$ can be represented by a MPO on the boundary $\partial V$, 
built from copies of a single local tensor $M$. Assuming now that we already have this property for a 
single site $e$, there must exist a pseudo-inverse $\peps^{+}$ such that $\peps^{+} \peps= \openone_{S_e}$ 
for some MPO projector $\openone_{S_e}$.

Throughout the remainder of this Letter we make use of standard tensor diagram notation, depicting each 
tensor as a point (or shape) with a leg emerging for each vector space it acts upon, and where a leg joining two 
tensors implies contraction of the associated indices.

By applying the pseudoinverse $\peps^{+}$ to the physical leg of PEPS tensor $\peps$ we obtain a one-to-one 
correspondence between virtual and physical degrees of freedom within the subspace $S$. Pictorially,
\begin{equation}\label{MPOinj}
\begin{centering}\vcenter{\hbox{
\includegraphics[width=0.5\linewidth]{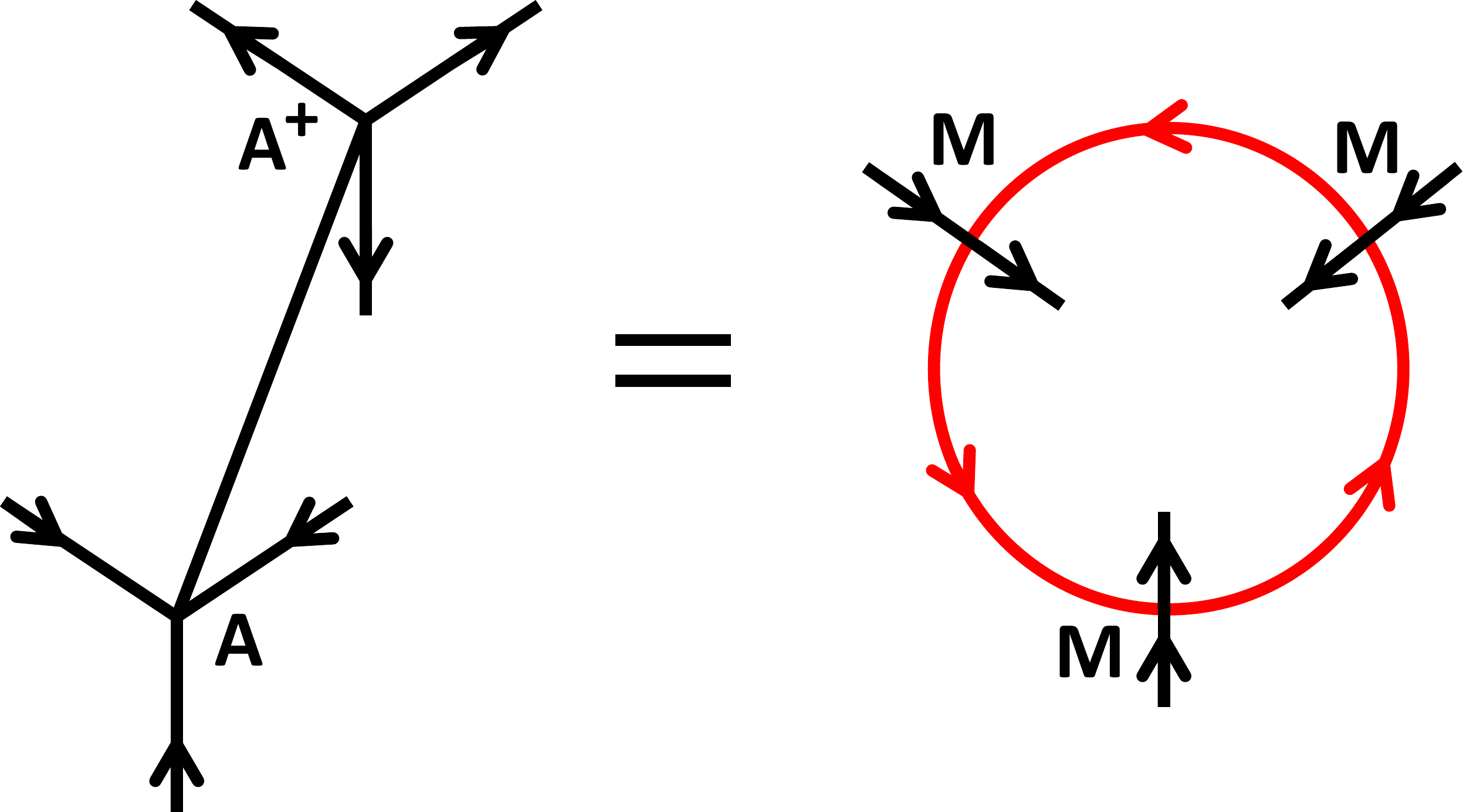}
}}\end{centering},
\end{equation}
where we make the convention that a PEPS tensor is associated to each intersection of black lines, and a MPO 
tensor, denoted by $M$ on the right hand side of Eq.~\eqref{MPOinj} where
$M: \mathbb{C}^D \otimes \mathbb{C}^m \rightarrow \mathbb{C}^D \otimes \mathbb{C}^m$, 
to each intersection of a black and red line (denoting $\mathbb{C}^m$). 

\simplesection{Algebraic Rules for Topological Order}
The main aim of MPO injectivity is to characterize
gapped quantum phases at zero temperature using the PEPS framework, including all 
models that give rise to a topological degeneracy of the ground state. 
In light of the now well established understanding of two dimensional topological models \cite{Kitaev03} 
and topology dependent ground state subspaces \cite{2012arXiv1207.1671H} we seek to describe general 
topological phases using string-like operators that can be arranged along noncontractable loops of the surface.
The definition of MPO injectivity places the physical degrees of freedom in one-to-one correspondence with
the virtual degrees of freedom in a certain subspace, allowing us to import the properties of string operators and ground states 
to MPOs on the virtual level. This approach is advantageous for the description of models away from a renormalization 
fixed point as we can essentially treat string operators on the virtual level as if we were at a fixed point, 
avoiding the need to broaden the string operators over some correlation length.

Except for at open endpoints, the MPOs must be free to move through the lattice to ensure that they are locally unobservable. 
This requirement can be satisfied locally by imposing the following condition.

\textbf{Pulling-through condition:} MPOs are free to pass
through the PEPS tensor $\peps$ on the virtual level
\begin{equation}\label{Pullingthrough}
\begin{centering}\vcenter{\hbox{
\includegraphics[width=0.5\linewidth]{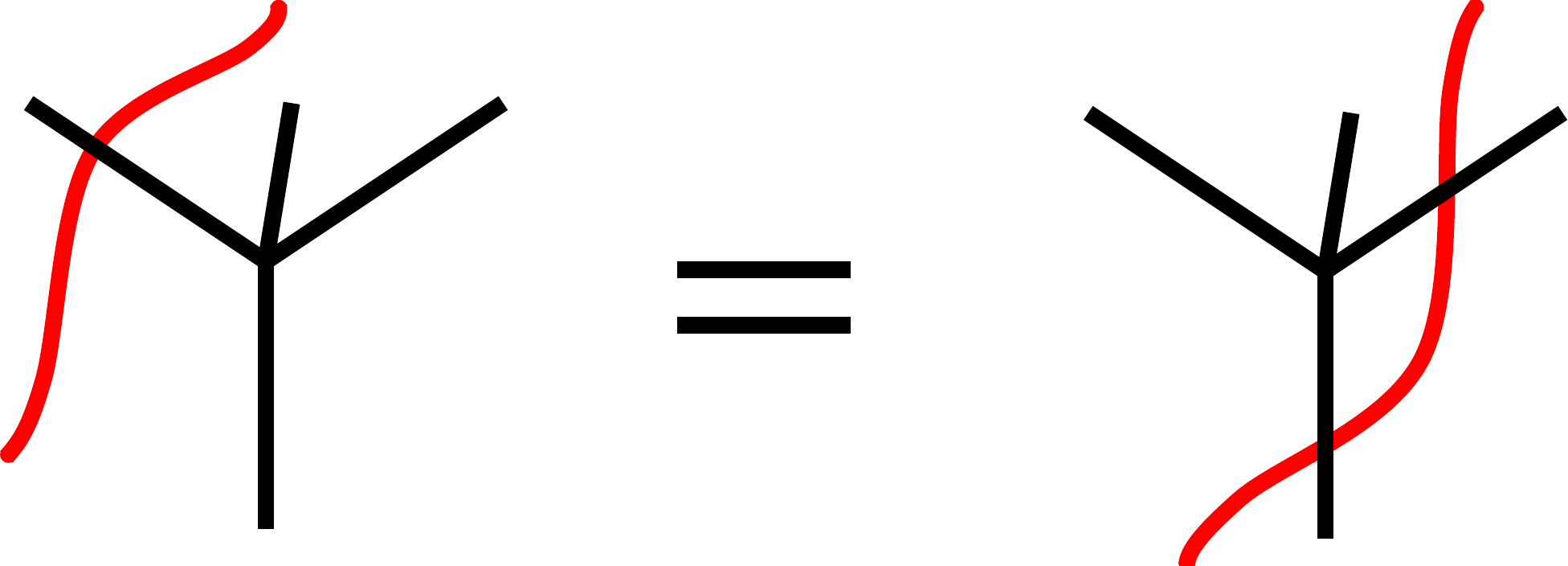}
}}\end{centering}.
\end{equation}
By applying the pseudo inverse $\peps^{+}$ and Eq.~\eqref{MPOinj}, we can express this condition purely 
in terms of MPO tensors. We also require the following condition for a trivial MPO loop
\begin{equation}\label{null}
\begin{centering}\vcenter{\hbox{
\includegraphics[width=0.35\linewidth]{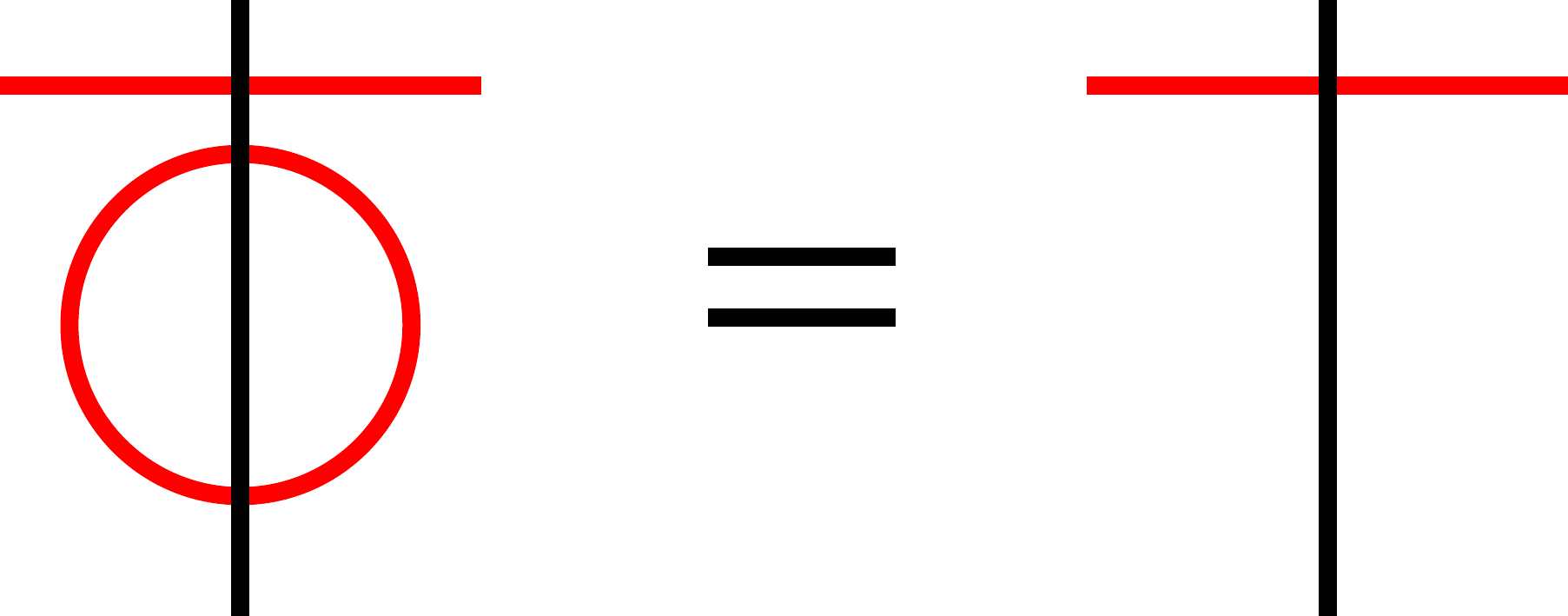}
}}\end{centering},
\end{equation}
which guarantees that any closed MPO is a projector, as required by the definition of MPO injectivity. 
This can easily be seen by taking two elementary loops of MPOs, pulling one through the other onto a single 
leg (using \eqref{MPOinj} and \eqref{Pullingthrough}) and then applying \eqref{null}.

\textbf{Generalized inverse:} The final condition required for a consistent definition of the invariant subspace 
$S$ on arbitrary regions of the lattice is the existence of a tensor $X$,
\begin{equation}\label{generalizedinverse}
\begin{centering}\vcenter{\hbox{
\includegraphics[width=0.3\linewidth]{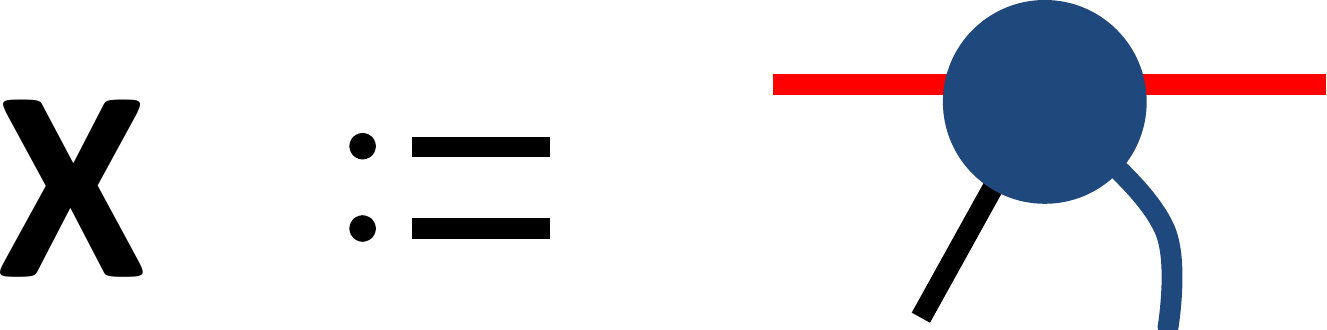}
}}\end{centering},
\end{equation}
for which 
\begin{equation}\label{Inverse}
\begin{centering}\vcenter{\hbox{
\includegraphics[width=0.5\linewidth]{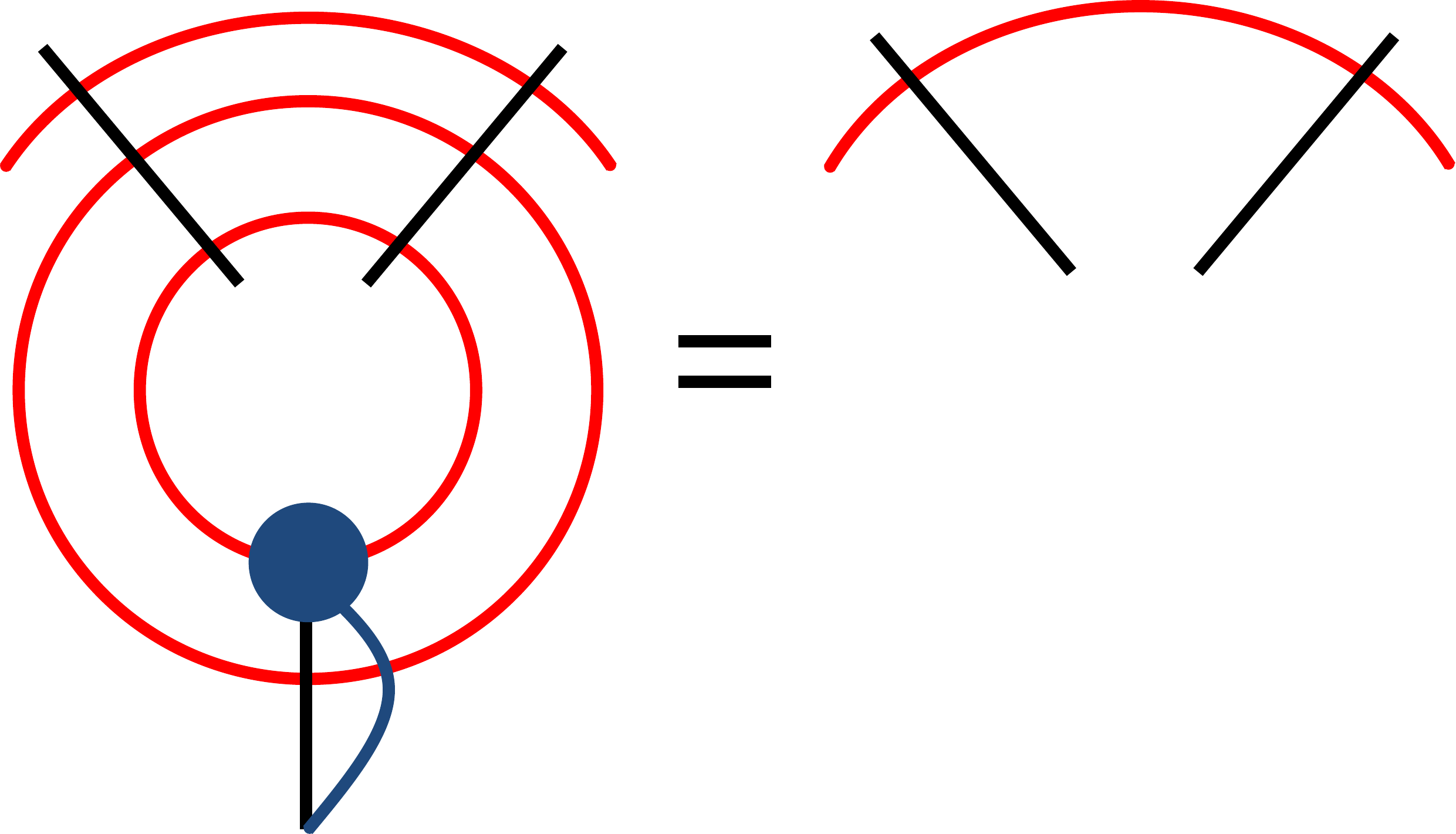}
}}\end{centering}
\end{equation}
holds. Note that there is no tensor associated to the intersection of blue and red lines. Together with the 
pulling through condition [Eq.~\eqref{Pullingthrough}], Eq.~\eqref{Inverse} implies that the MPO injectivity 
condition is stable when multiple PEPS tensors are concatenated. It further implies that the ground space 
of the parent Hamiltonian on a contiguous region (which is given by a sum of local terms) is frustration free 
and is spanned by the concatenated PEPS tensors on the region with arbitrary states on the virtual boundary, 
which is known as the intersection property. Proofs of the stability under concatenation and the intersection 
property are provided in the supplementary material.

We conjecture that in two spatial dimensions all gapped, topologically ordered ground states admitting a PEPS description 
can be constructed from MPO injective [Eq.~\eqref{MPOinj}] tensors, with the MPO arising as a solution of
Eq.~\eqref{Pullingthrough} and Eq.~\eqref{null} for which there exists a tensor $X$ satisfying 
Eq.~\eqref{Inverse}.

\simplesection{Ground state subspace} As a result of the algebraic rules, the ground 
state subspace of a MPO injective PEPS parent Hamiltonian is spanned by a finite number of states. If the 
tensor network is closed on a torus, we find (see supplementary material) that the ground 
state subspace is spanned by states obtained from tensors $Q$ that satisfy
\begin{equation}\label{Ground state space}
\begin{centering}\vcenter{\hbox{
\includegraphics[width=0.85\linewidth]{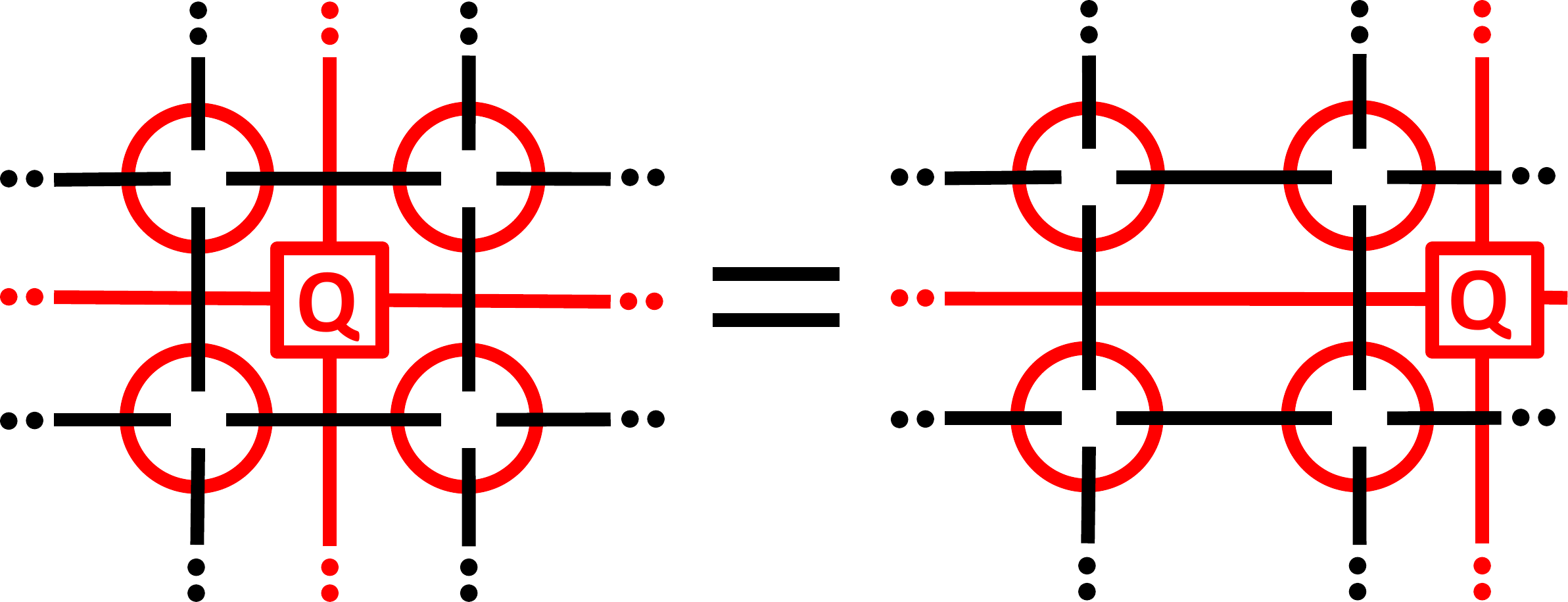}
}}\end{centering},
\end{equation}
where the dots indicate periodic boundary conditions, and the equivalent equation for the vertical direction. This ensures that the resultant physical ground states are translation invariant and the location of the closure is of no significance.
The ground state degeneracy on the torus is then given by the number of linearly independent physical 
states arising from the solution of Eq.~\eqref{Ground state space} that remain distinct and normalizable in the thermodynamic limit (see appendix).

As this formalism was set up in order to characterize all gapped quantum phases, it naturally also includes states with discrete symmetry breaking (i.e. cat states). In that case, the different ground states can have local order parameters and the degeneracy will be independent of the topology. Throughout the remainder of this letter, we focus on the more interesting case of topological phases. To determine whether any of the resulting degeneracy is truly topological in nature, one must compare the ground state degeneracy of the tensor network on the topological 
manifold of interest to  that arising on a topologically trivial manifold (such as the sphere). This provides a deterministic recipe for checking, for any given model, the topological dependency of the ground state subspace.

\simplesection{Identifying the topological order} 
Let us now discuss how to identify the topological order in MPO injective
PEPS.  Since MPO injectivity is stable under concatenation, for any
contiguous region the virtual indices at the boundary are supported on
the invariant subspace of the MPO.  This is on the one hand reflected in the low-energy excitations at the
edge, which are in one-to-one correspondence with admissible boundary
conditions. The edge dynamics are thus restricted to the invariant
subspace of the MPO~\cite{yang:peps-edgetheories}, which provides topological protection to the
edge and allows to infer the structure of the MPO from the edge physics.  At the same time, it is reflected in the
entanglement spectrum and the corresponding entanglement
Hamiltonian~\cite{li:es-qhe-sphere}. The entanglement spectrum is also restricted to the invariant subspace, and therefore, the entanglement
Hamiltonian contains a universal term with infinite strength which
restricts the system to be in the invariant
subspace~\cite{schuch:topo-top}.  A consequence of this restriction
(together with the MPO injectivity) is that the number of non-zero
eigenvalues is equal to the dimension of the invariant subspace of the
MPO, which gives rise to a topological correction to the zero R\'enyi
entropy. In the case of RG fixed points, the
correction should not depend on the R\'enyi index (as has been shown for
string-net models~\cite{topologicalrenyi}) which implies a corresponding
topological correction to the entanglement
entropy~\cite{KitaevPreskill,levin:topological-entropy}.

The topological correction does not fully characterize the topological
phase.  To this end, one must
obtain the modular $S$ and $T$ matrices which contain all the relevant
information about the topological excitations such as the mutual and self
braiding statistics~\cite{Haah14}. The fusion rules of the topological
excitations can be obtained from the $S$ matrix via the Verlinde formula.
An advantage of the MPO formalism is that it allows for an unambiguous
definition of modular transformations on the ground states of a lattice
system on a torus, obtained by solving Eq.~\eqref{Ground state space}.
The $90^{\circ}$ rotation can be performed directly on the ground state
tensors $Q_i$ defined in Eq.~\eqref{Ground state space}.  The Dehn twist,
on the other hand, corresponds to increasing the winding number of the MPO
along the twisting direction by one. If one uses $\peps^{+}\peps$ for the
PEPS tensors then the overlap matrix of the original ground states with
the rotated (twisted) ground states will only contain universal
information and therefore correspond to the $S$ ($T$) matrix~\cite{modular1,modular2,modular3}.

The solutions of Eq.~\eqref{Ground state space} will in general
not correspond to the minimally entangled states (MES), i.e. the states
that have the physical interpretation of being threaded with a definite
anyon flux through one of the holes of the torus \cite{modular1}.
Thus, one first has to find a unitary basis transformation of the ground
state subspace (corresponding to a basis transformation of the tensors
$Q$) that makes $T$ diagonal and $S$ symmetric \cite{modular4} to be able
to read off the topological properties of the excitations.
Note that by wrapping the MPO around the torus in one direction, we can
always construct a state with a topological flux corresponding to some Abelian
anyon threaded through the hole in the orthogonal direction, since these are 
states with maximal topological entropy $2\gamma$ (while for general topological 
fluxes, the correction is $2\gamma - \log(d_i^2)$ with $d_i$ the 
quantum dimension). In the case that all anyons are non-Abelian 
this MES clearly corresponds to the one with a trivial flux. But since this
construction works for any anyon theory it is very likely that 
it will always lead to the MES with a trivial flux.

\simplesection{Example: String-net models} In this section, we show that the set of models described by 
MPO injective PEPS contains all string-net models~\cite{LevinWen05}---the largest set of many-body 
bosonic lattice models exhibiting topological order available in the literature. 
Details about the PEPS description of string-net models \cite{GuLevinSwingleWen09, BuerschaperAguadoVidal09} 
are provided in the supplementary material. 
We now proceed to show that for the tensor network description of string-net condensed states, both 
MPO injectivity [Eq.~\eqref{MPOinj}] and the pulling through condition [Eq.~\eqref{Pullingthrough}] are implied 
by the pentagon equation for the $F$-symbols. For simplicity, we demonstrate this in the multiplicity free 
case and note that the direct generalization applies to all string-net models. We start from the definition of the string-net PEPS tensor
\begin{equation}\label{string-net tensor}
\begin{centering}\vcenter{\hbox{
\includegraphics[width=0.5\linewidth]{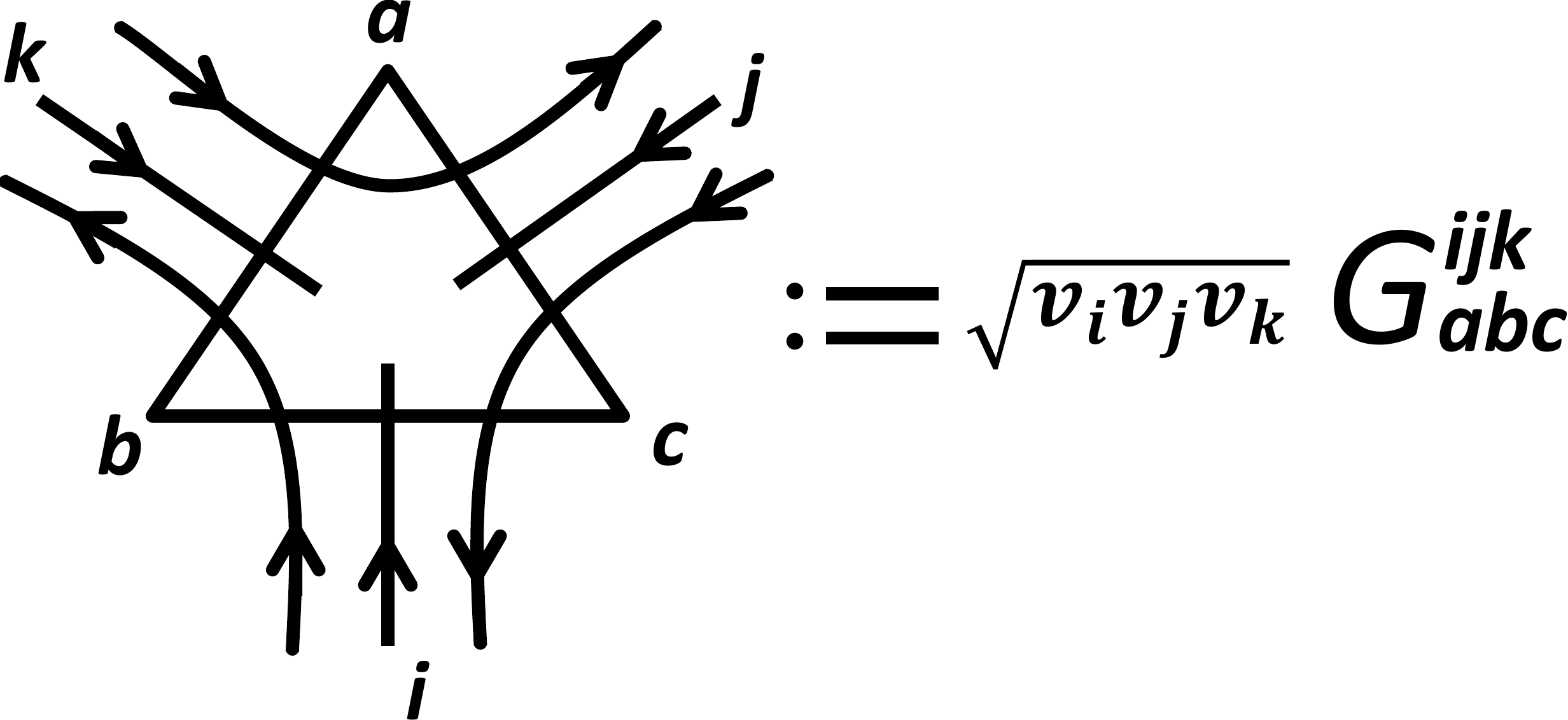}
}}\end{centering},
\end{equation}
where the $i,j,k$ legs are copied to the physical level, and the MPO tensor
\begin{equation}\label{string-net MPO}
\begin{centering}\vcenter{\hbox{
\includegraphics[width=0.5\linewidth]{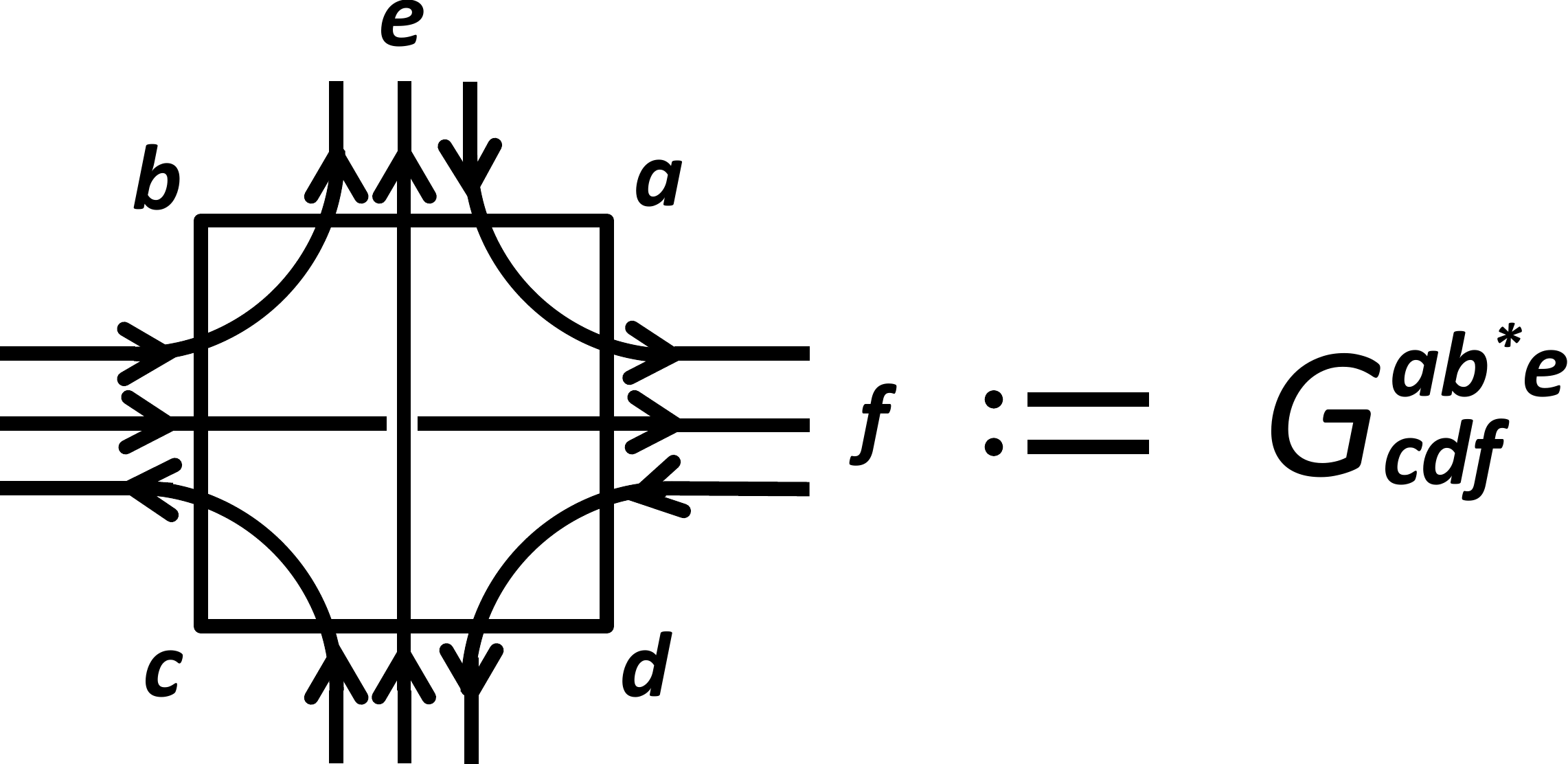}
}}\end{centering},
\end{equation}
note that we explicitly depict all tensors as 2D shapes for the string-net PEPS.
The $G$-symbol is a symmetrized version of the $F$-symbol, defined in Eq.~\eqref{gsymbol}, that 
is invariant under simultaneous cyclic permutation of the upper and lower indices.
These diagrams use the convention that a pair of tensor legs $i$, $i'$, that are connected through the body 
of a tensor corresponds to a Kronecker delta on the associated indices, i.e.\ $T_{\{j\},i,i'}=\tilde{T}_{\{j\},i} \delta_{i,i'}$; 
we therefore use a single label in the pictures. As a final convention, we always associate a multiplicative 
factor of the quantum dimension $d_{\lambda}$ to each term in a sum over any index $\lambda$ 
(appearing as a closed loop in the diagrams below)~\footnote{This convention can be implemented locally by adding multiplicative factors to the string-net PEPS~\eqref{string-net tensor} and MPO~\eqref{string-net MPO} tensors such that every closed loop of $\lambda$ gets a factor of quantum dimension $d_\lambda$. We attach a factor of $d^{(1 - \alpha / \pi)/2}_\lambda$ to every bending line of the string-net PEPS and MPO tensors, where $\alpha$ is the bending angle in radians. Then, for any closed loops with $n$ bending points, i.e., polygons with $n$ edges, we get $d^{(n - \alpha'/\pi)/2}_\lambda=d_\lambda$ because $\alpha'$, the total interior angle of $n$-polygon, is equal to $(n-2)\pi$ \label{note} }. 

We are now in a position to define the pseudo-inverse and demonstrate that
it satisfies equation \eqref{MPOinj} for the string-net PEPS and MPO tensors defined in  Eq.~\eqref{string-net tensor} 
and Eq.~\eqref{string-net MPO}. We then move on to show that the
pulling through condition in Eq.~\eqref{Pullingthrough} also holds for these PEPS and MPO tensors.
We point out that both of these identities are implied by the
pentagon equation~\eqref{pentagoneqn}, which appears as a compatibility
condition for the $F$-symbols~\cite{LevinWen05} and is thus guaranteed to be true for any
string-net model.

The pseudo-inverse is given by
\begin{equation}\label{pseudo-inverse}
\begin{centering}\vcenter{\hbox{
\includegraphics[width=0.5\linewidth]{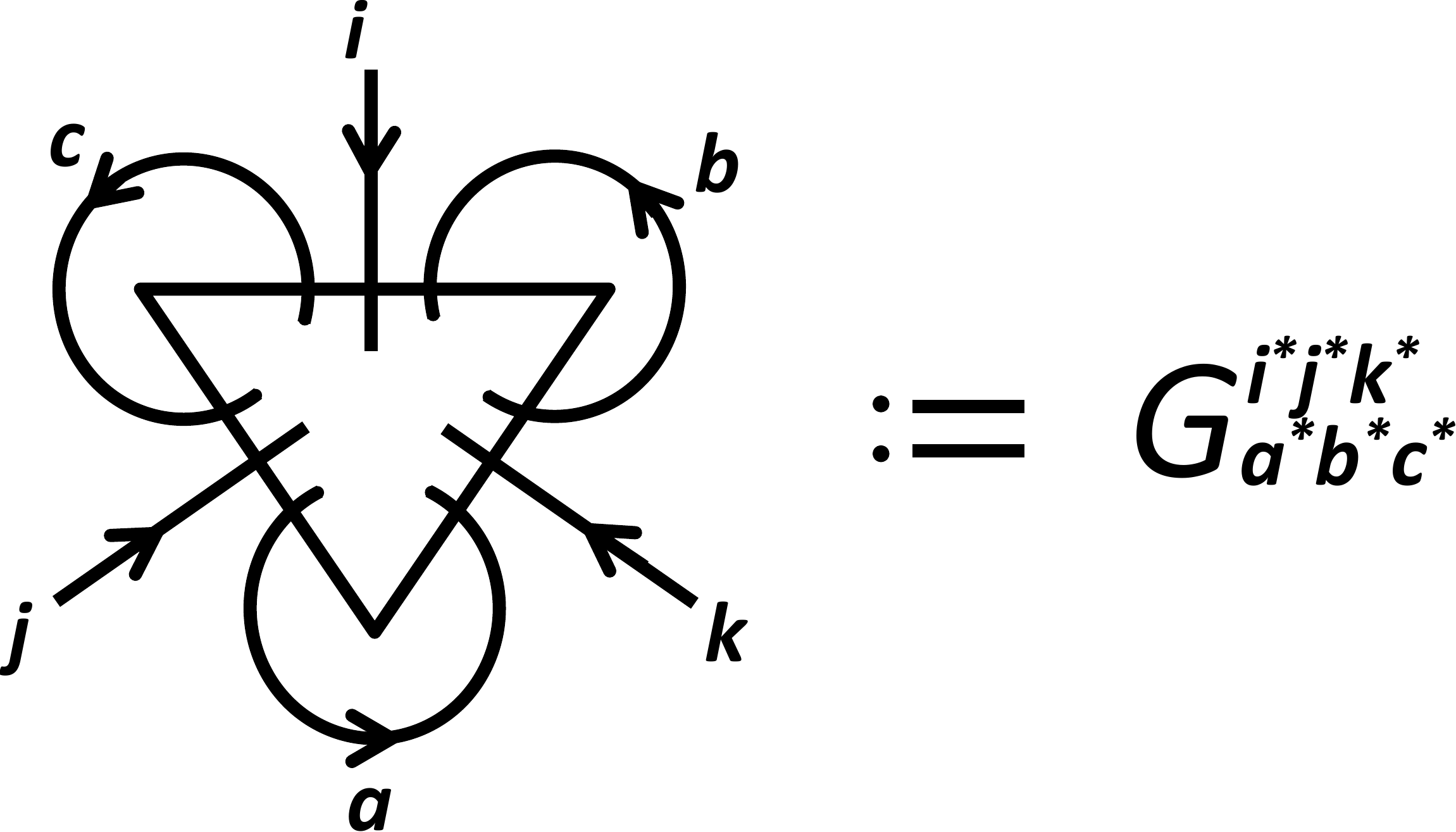}
}}\end{centering},
\end{equation}
whence MPO injectivity \eqref{MPOinj} follows from
\begin{equation}\label{string-net MPOinj}
\begin{centering}\vcenter{\hbox{
\includegraphics[width=0.84\linewidth]{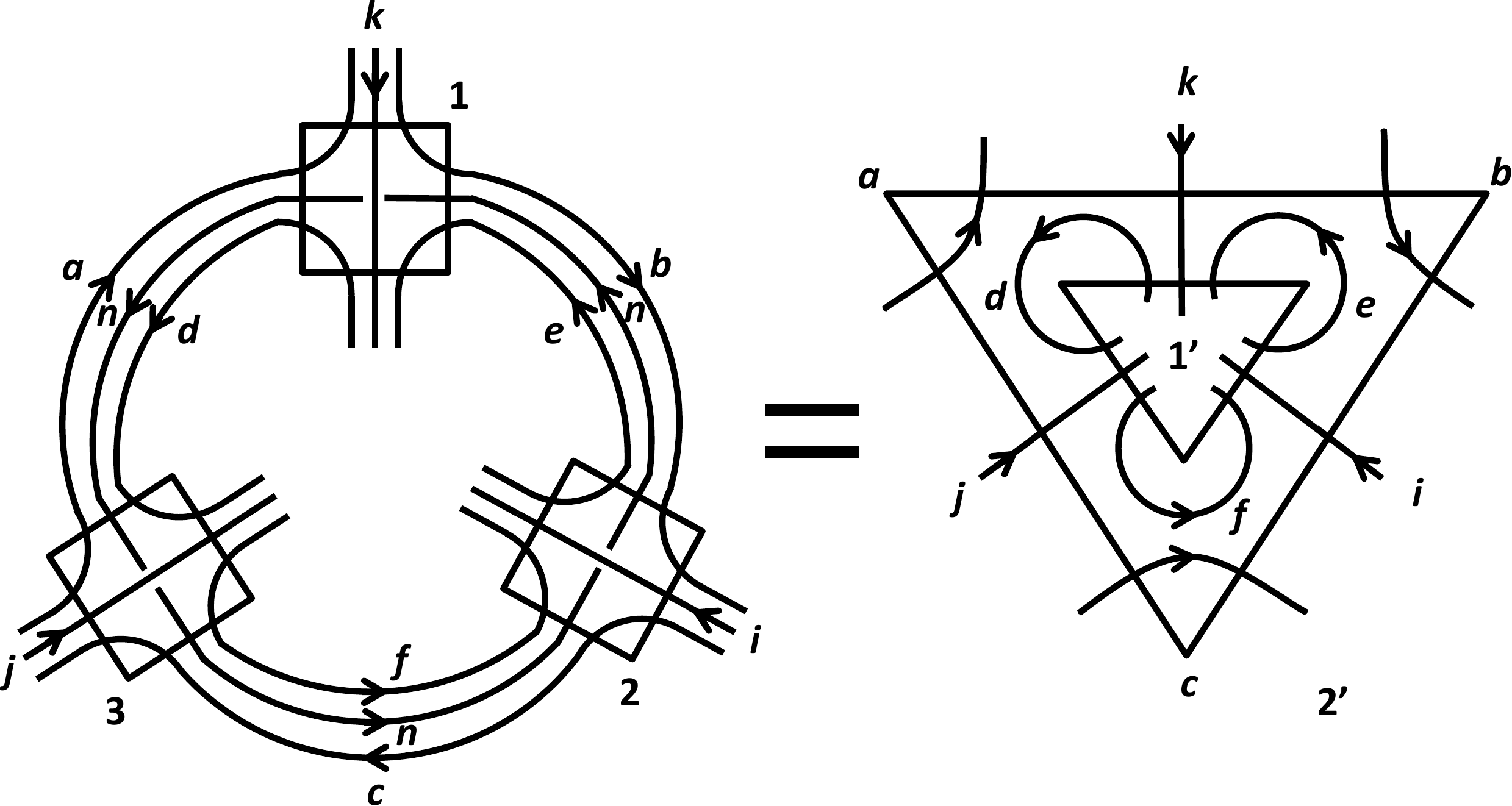}
}}\end{centering},
\end{equation}
(note the sum over $n$ and the associated quantum dimension $d_n$) which is the pentagon equation [Eq.~\eqref{pentagoneqn}] as shown by 
\begin{equation}\label{pentagon MPOinj}
\begin{centering}\vcenter{\hbox{
\includegraphics[width=0.8\linewidth]{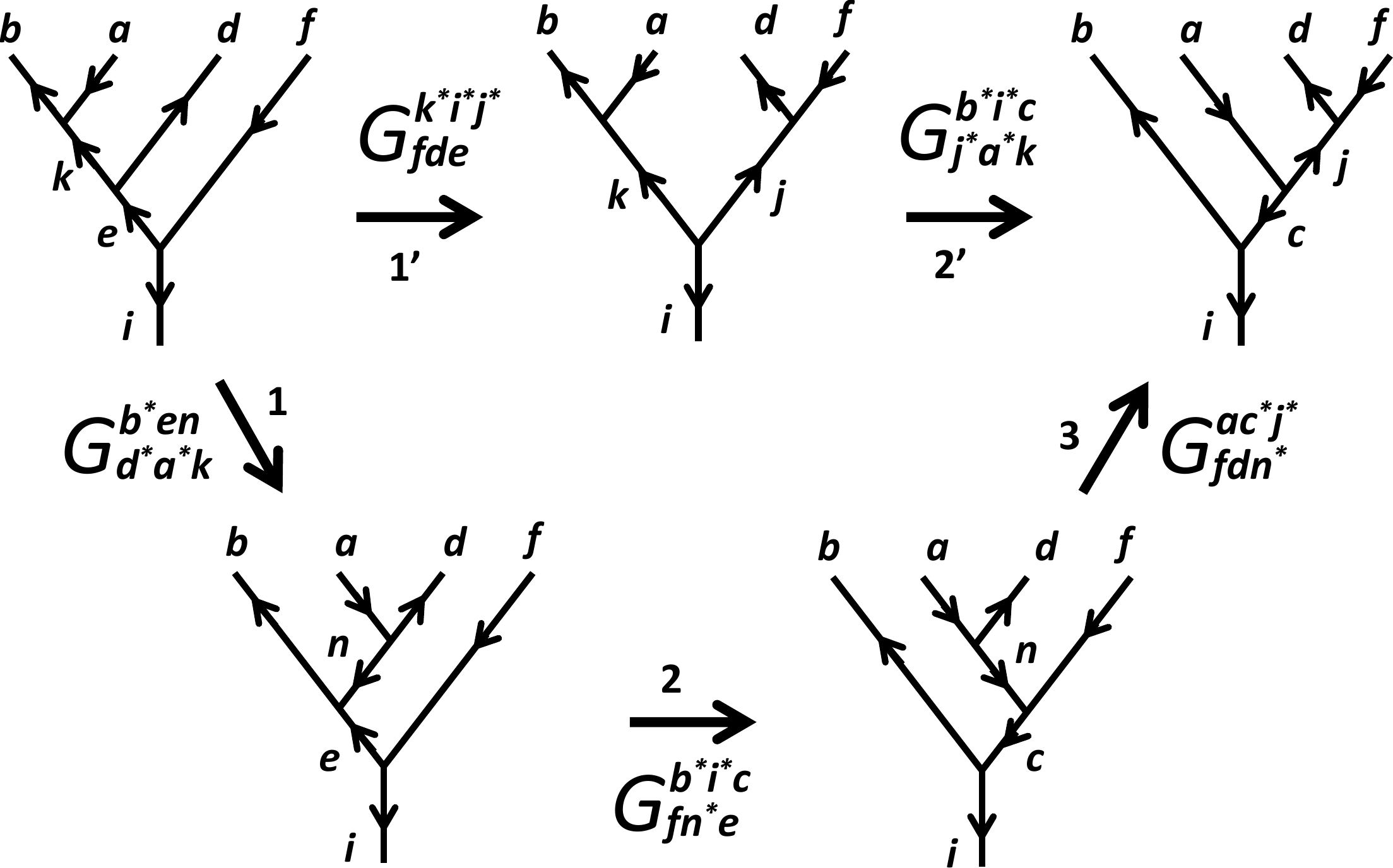}
}}\end{centering}.
\end{equation}
The number labeling each move in the above diagram indicates which of the tensors in Eq.~\eqref{string-net MPOinj} 
the move corresponds to. The $G$-symbol next to each move in Eq.~\eqref{pentagon MPOinj} comes from the definition in Eq.~\eqref{G picture} and is equal to the corresponding tensor in  Eq.~\eqref{string-net MPOinj}, which can be seen by employing several identities of the $G$-symbol [Eq.~\eqref{Symmetry F}].

The following is the pulling through condition for the string-net PEPS and MPO tensors
\begin{equation}\label{string net - pulling through}
\begin{centering}\vcenter{\hbox{
\includegraphics[width=0.84\linewidth]{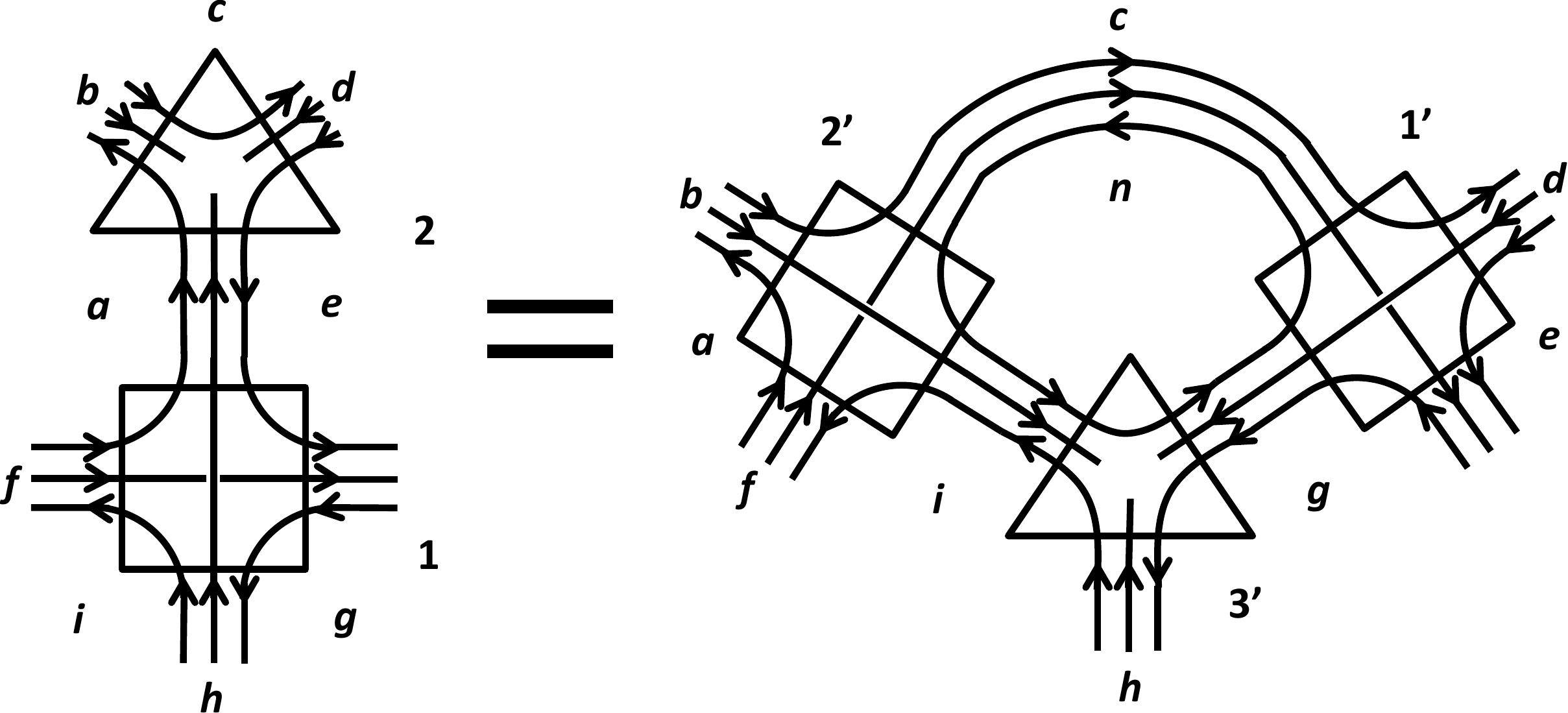}
}}\end{centering},
\end{equation} 
which again is the pentagon equation [Eq.~\ref{pentagoneqn}]
\begin{equation}\label{pentagon pulling through}
\begin{centering}\vcenter{\hbox{
\includegraphics[width=0.8\linewidth]{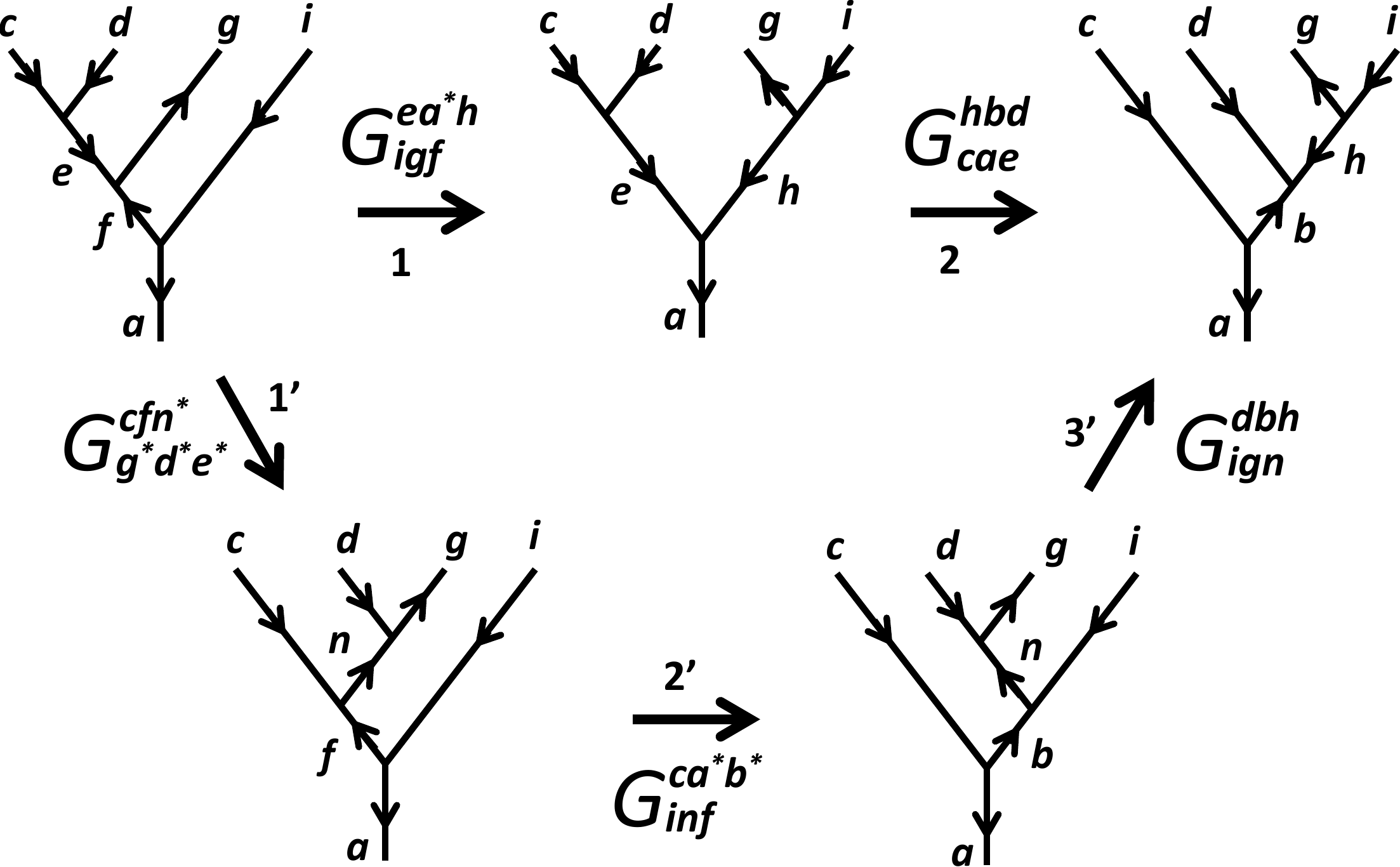}
}}\end{centering}.
\end{equation} 
The loop condition [Eq.~\eqref{null}] for string-net PEPS follows from the unitarity of the $F$-symbols as a basis transformation. The existence of a generalized inverse $X\in
\mathbb{C}^m\otimes\mathbb{C}^m\otimes\mathbb{C}^D\otimes\mathbb{C}^D$, 
for which Eq.~\eqref{generalizedinverse} holds, follows from Eq.~\eqref{string-net MPOinj} 
(i.e. the pentagon equation) and the unitarity of the $F$-symbols.

One can readily verify that a closed string-net MPO constructed from the tensors in Eq.~\eqref{string-net MPO}, with a weighting of the normalized quantum dimension $d_i/D^2$ associated to the internal index forming a closed loop, is a projector for any length. Since the MPO is a projector its rank is easily obtained by calculating the trace. By examining the behavior of this rank as the length increases it is clear that the topological entanglement entropy is $\log(D^2)$, originating from the overall normalization factor $D^{-2}$.

\simplesection{Conclusions}
In this Letter, we have presented a general framework for the
characterization of topological phases in two dimensions using the PEPS
formalism. The key ingredient is a generalized notion
of injectivity, in which central object is a MPO fulfilling a fundamental
\emph{pulling through} condition.  Both $\mathsf{G}$
injectivity~\cite{SchuchCiracGarcia10} and twisted
injectivity~\cite{Buerschaper14} turn out to be special cases of MPO
injectivity, which can be verified directly by constructing the relevant
MPOs.  As a very general example, we illustrated that all string-net
models satisfy our axioms by explicitly constructing the appropriate MPOs
and elucidating the correspondence between the pentagon equation and our
pulling-through condition.  


The characterization of topological order in terms of MPOs opens up the
possibility of classifying topological phases via their MPOs, similar to
results on 2D symmetry-protected topological (SPT) phases~\cite{ChenLiuWen11}. In fact, for string-net
models with abelian group elements as local degrees of freedom on edges and
with group multiplication as the trivalent vertex constraints, the pentagon equation
reduces to a $3$-cocycle condition. 
This close connection between the boundary symmetry MPOs of SPT states
and the virtual gauge symmetry MPOs of intrinsic topological states is
explicitly described by a gauging duality in the PEPS picture~\cite{HaegemanAcoleyenSchuchCiracVerstraete14}.


The framework set forth in this paper can be easily generalized to
fermionic PEPS~\cite{fpeps}, as well as to higher dimensional systems by
replacing MPOs with their higher dimensional generalization, Projected
Entangled Pair Operators (PEPOs); it thus provides a systematic way to
understand both topological phases of interacting fermions and exotic
topological order in three dimensions such as the Haah
code~\cite{haah:haah-code}.


A natural question is whether our framework contains topological phases
outside the string-net picture.  Since the excitations of the doubled
phases described by string-nets all have a Lagrangian subgroup we know
that their edge modes can be gapped out~\cite{levin}. Thus, to obtain
phases outside string-nets we need to look at models with protected
gapless edge modes (or models which do not correspond to a TQFT). Given
the close connection between edge physics and the MPO, this amounts to
understanding which MPOs give rise to protected gapless edge modes;
indeed, in the recently discovered chiral fPEPS~\cite{wahl:chiral-fPEPS},
fermionic MPOs satisfying a pulling-through condition have been 
identified~\cite{wahl:chiral-fPEPS-edge}.

Finally, an equation closely related to the pulling through condition
[Eq.~\eqref{Pullingthrough}] could yield an algorithm to bring 2D PEPS
into a normal form that facilitates the calculation of physical
observables. Intuitively, this is because once the algorithm has converged
we find an MPO that approximates the transfer matrix of the model. Hence,
contracting the whole PEPS with a physical observable reduces to
contracting the PEPS in a local region around the observable and using a
MPO to approximate the boundary.  We leave these directions to future
work.

\bibliography{MPO-bib}

\appendix
\section{Supplementary Material}

\textbf{Stability under concatenation:} We now show that the conditions we have placed on the MPO tensors [Eq.~\eqref{Pullingthrough}, Eq.~\eqref{null} and Eq.~\eqref{generalizedinverse}] ensure that the projector $\openone_ {S_V}$ onto the virtual subspace $S_V$, on which the PEPS map acts injectively, is represented by a MPO for any simply connected region of the lattice with nontrivial boundary. The pseudoinverse of the PEPS map on a larger region can be constructed by first applying a pseudoinverse to each site and then using the pulling through condition and applying the generalized inverse $X$, as represented by the moves in the following diagrams
\begin{equation}\label{concatenation1}
\begin{centering}\vcenter{\hbox{
\includegraphics[width=0.7\linewidth]{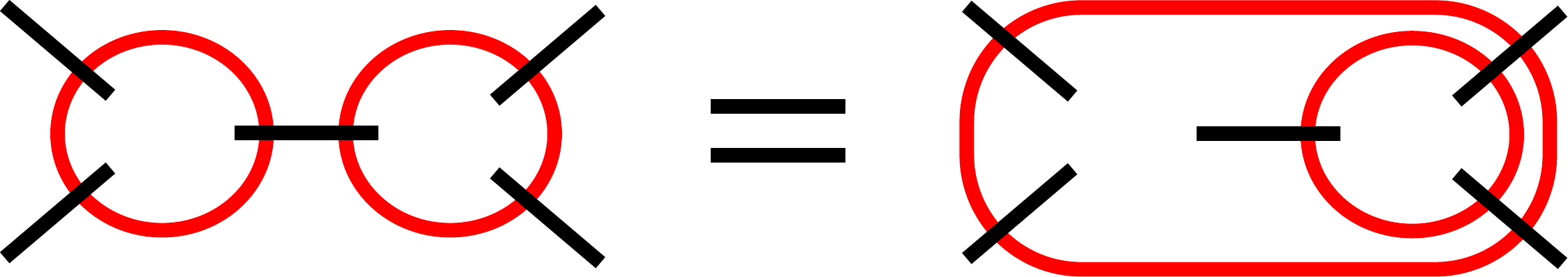}
}}\end{centering}
\end{equation} 
and then
\begin{equation}\label{concatenation2}
\begin{centering}\vcenter{\hbox{
\includegraphics[width=0.7\linewidth]{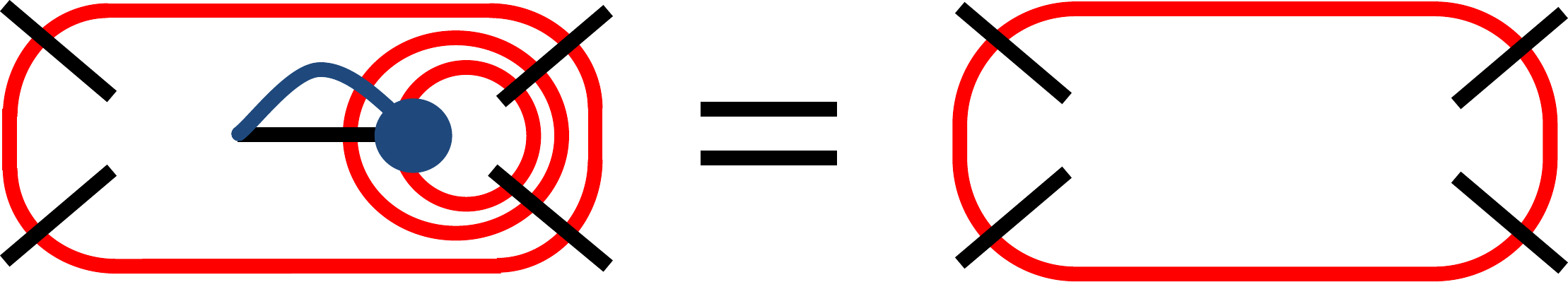}
}}\end{centering}.
\end{equation}
The same two moves can be used to inductively grow a region from $N$ to $N{+}1$ sites. The only complication arises when the injective region encloses an elementary plaquette, this involves growing the region onto a new site with two virtual bonds in common, which is possible using a slight variation of the above process.

\textbf{Renormalization group move:} The pulling through condition [Eq.~\eqref{Pullingthrough}] together with the generalized inverse [Eq.~\eqref{generalizedinverse}] yield natural maps (acting only upon the black indices) for the addition or removal of degrees of freedom to or from a MPO. 
One can construct a linear map which removes a single degree of freedom from an MPO as follows
\begin{equation}\label{rgunmove}
\begin{centering}\vcenter{\hbox{
\includegraphics[width=0.84\linewidth]{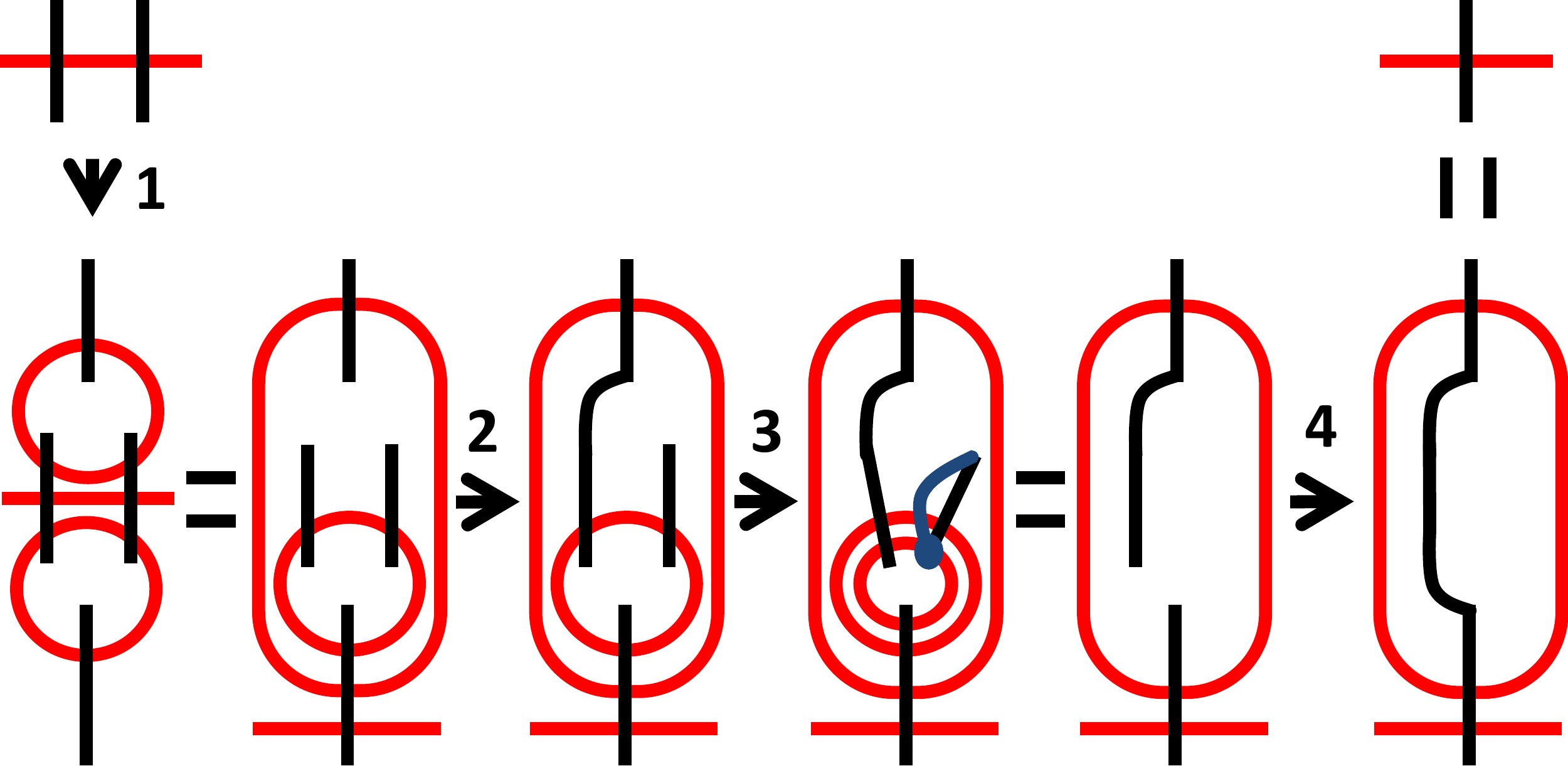}
}}\end{centering},
\end{equation}
where at step 1 we act with two MPO loops, at step 2 we contract two open indices, at step 3 we act with a generalized inverse, and at step 4 we again contract two open indices.
Adding a single degree of freedom can be done similarly
\begin{equation}\label{rgmove}
\begin{centering}\vcenter{\hbox{
\includegraphics[width=0.84\linewidth]{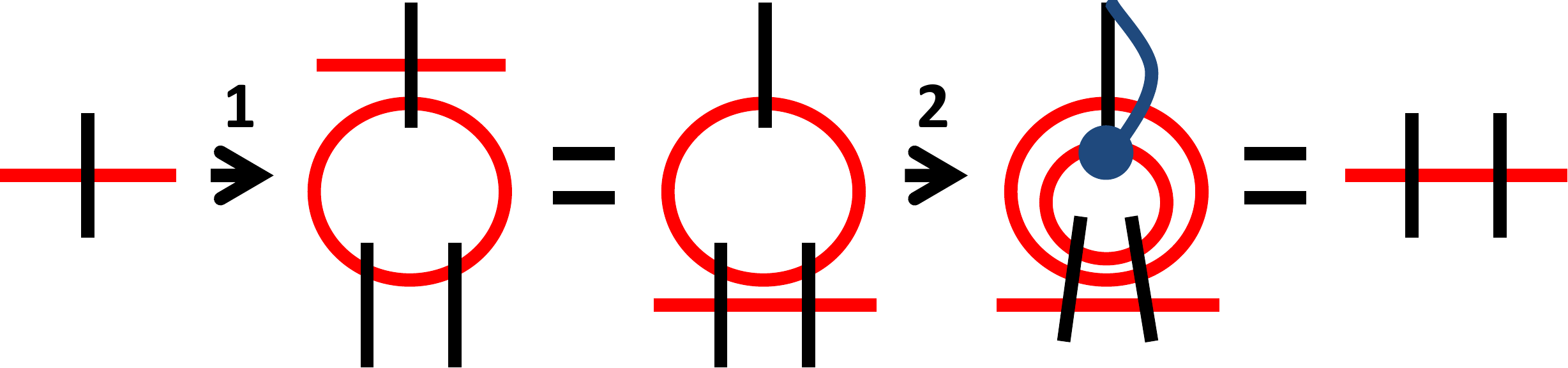}
}}\end{centering},
\end{equation}
where at step 1 we act with a single MPO loop and at step 2 we apply a generalized inverse.

These moves yield linear maps between MPO injective PEPS on lattices of different sizes and allow one to define a generalized inverse acting on any number of legs of an arbitrarily large MPO loop, and to show that the pulling through condition holds for arbitrarily large MPO loops. Both the coarse-grained generalized inverse and coarse-grained pulling through condition will be utilized in the following sections.

\textbf{Intersection:} In this section we show that the MPO injective PEPS parent Hamiltonian defined on 
any simply connected region of the lattice with nontrivial boundary is frustration free and, furthermore, that all 
states within the ground subspace are given by a unique tensor network representation built from the 
original PEPS tensors $\peps$ in the bulk with arbitrary tensors closing the network at the virtual boundary.

The parent Hamiltonian consists of a sum of $2\times2$ plaquette terms that each project locally onto the subspace spanned by the PEPS on that plaquette with arbitrary virtual boundary tensors. Here we consider the mutual ground state subspace of two neighbouring plaquette terms, any state within this subspace must be of the following form
\begin{equation}\label{i0}
\begin{centering}\vcenter{\hbox{
\includegraphics[width=0.84\linewidth]{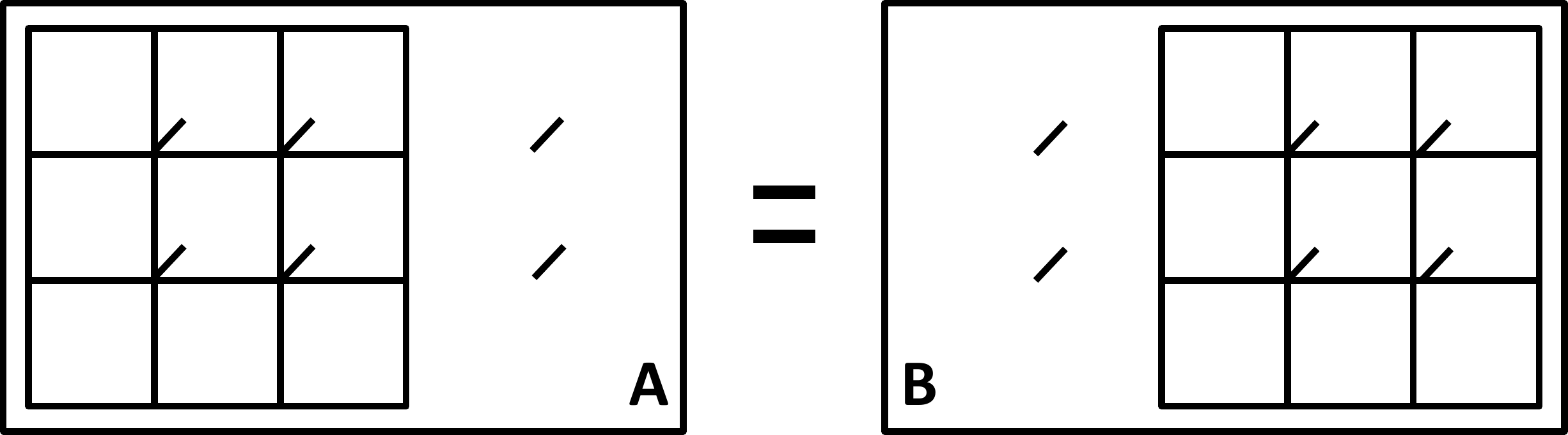}
}}\end{centering},
\end{equation}
for some boundary tensors $A$ and $B$, note that we are free to choose $B$ to be invariant under a loop of MPO on the virtual boundary.
By applying the pseudoinverse to all sites we find
\begin{equation}\label{i1}
\begin{centering}\vcenter{\hbox{
\includegraphics[width=0.84\linewidth]{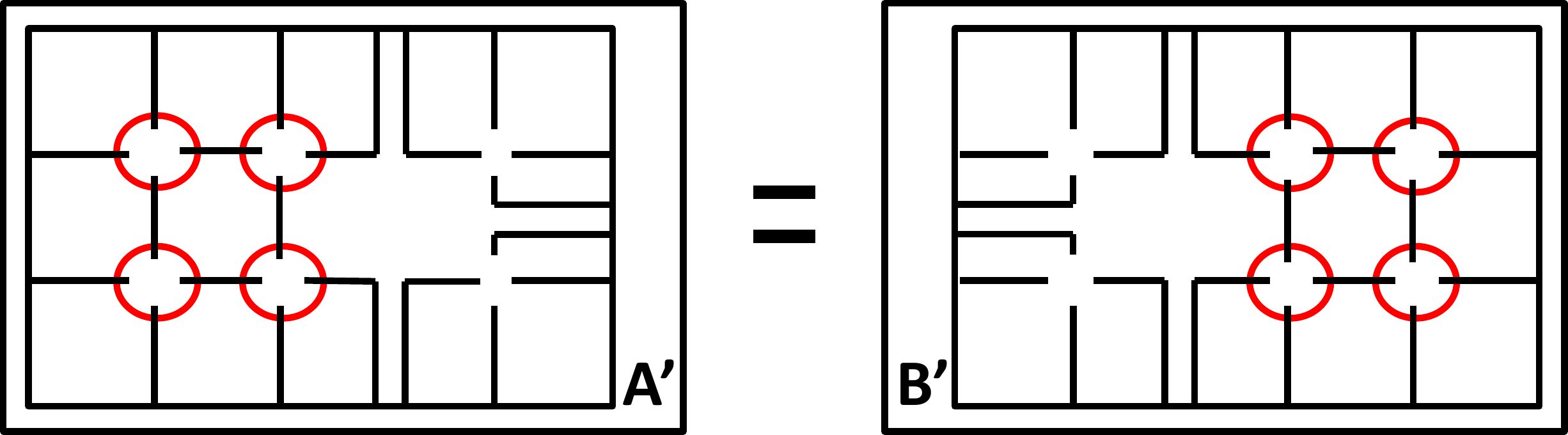}
}}\end{centering} ,
\end{equation}
which, after pulling through and applying a generalized inverse, leads to
\begin{equation}
\begin{centering}\vcenter{\hbox{
\includegraphics[width=0.84\linewidth]{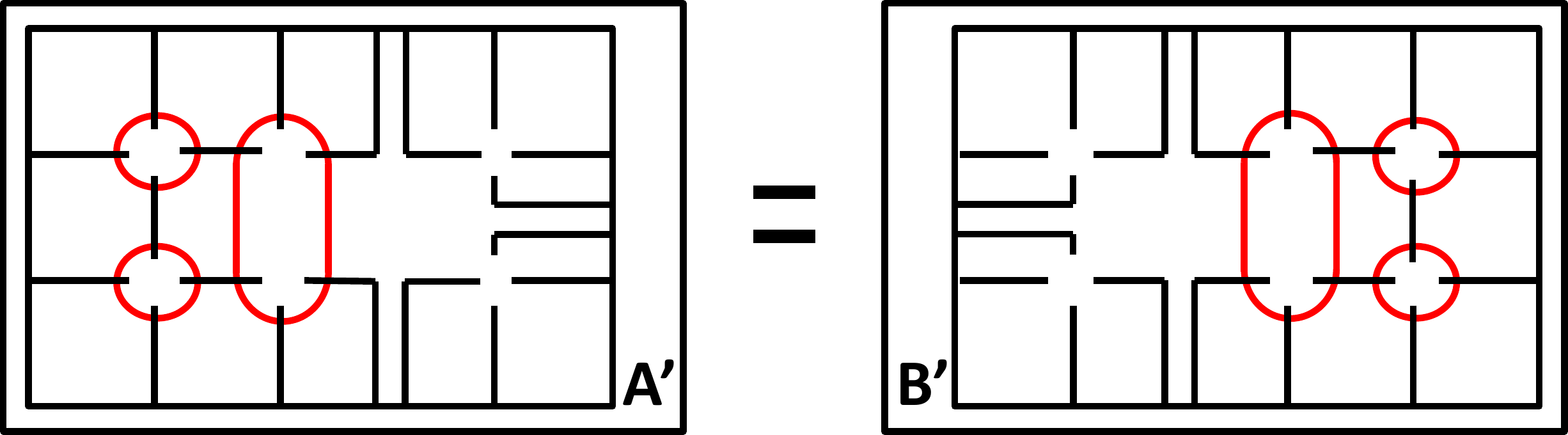}
}}\end{centering} .
\end{equation}
The application of another generalized inverse yields
\begin{equation}\label{i2}
\begin{centering}\vcenter{\hbox{
\includegraphics[width=0.84\linewidth]{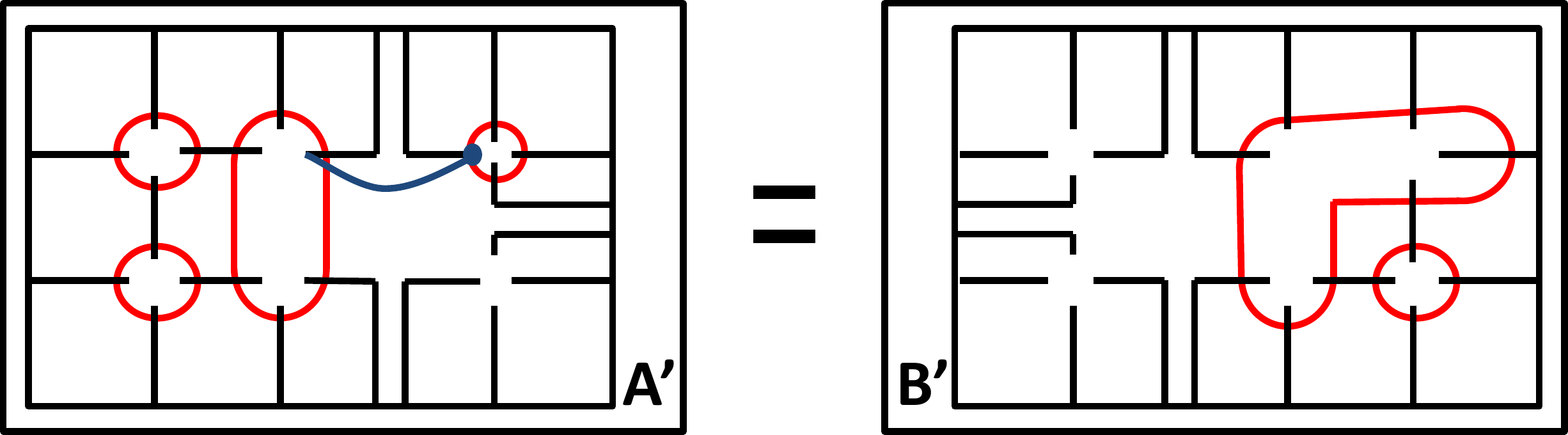}
}}\end{centering}
\end{equation}
and after applying a coarse-grained generalized inverse over two legs we find
\begin{equation}\label{i3}
\begin{centering}\vcenter{\hbox{
\includegraphics[width=0.84\linewidth]{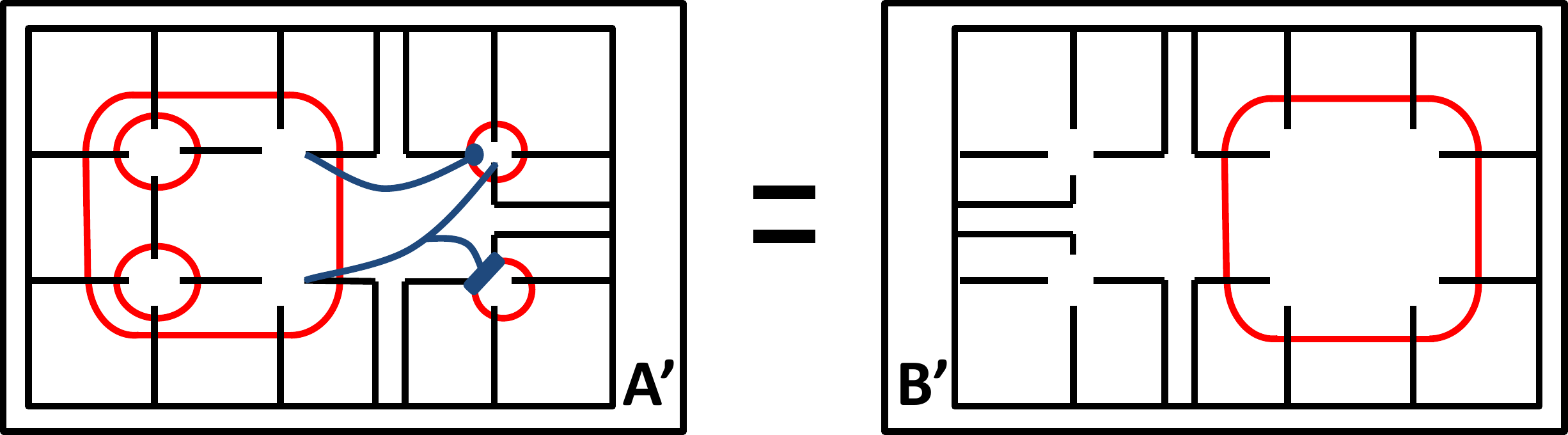}
}}\end{centering} .
\end{equation}
We define a new boundary tensor
\begin{equation}
\begin{centering}\vcenter{\hbox{
\includegraphics[width=0.84\linewidth]{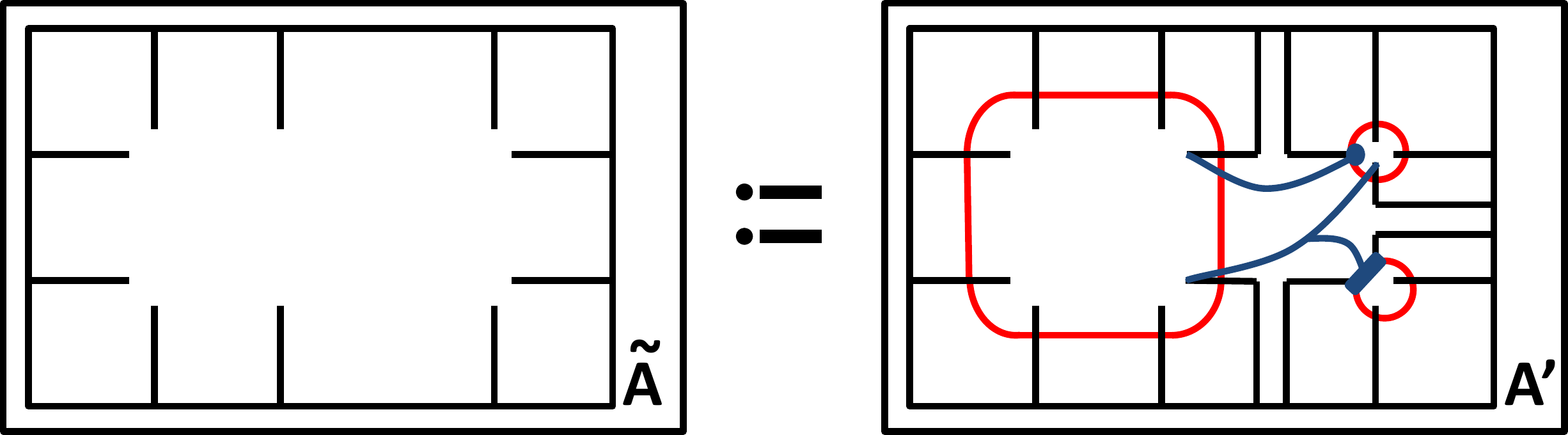}
}}\end{centering}
\end{equation}
and find that $B'$ must take the following particular form
\begin{equation}\label{i4}
\begin{centering}\vcenter{\hbox{
\includegraphics[width=0.84\linewidth]{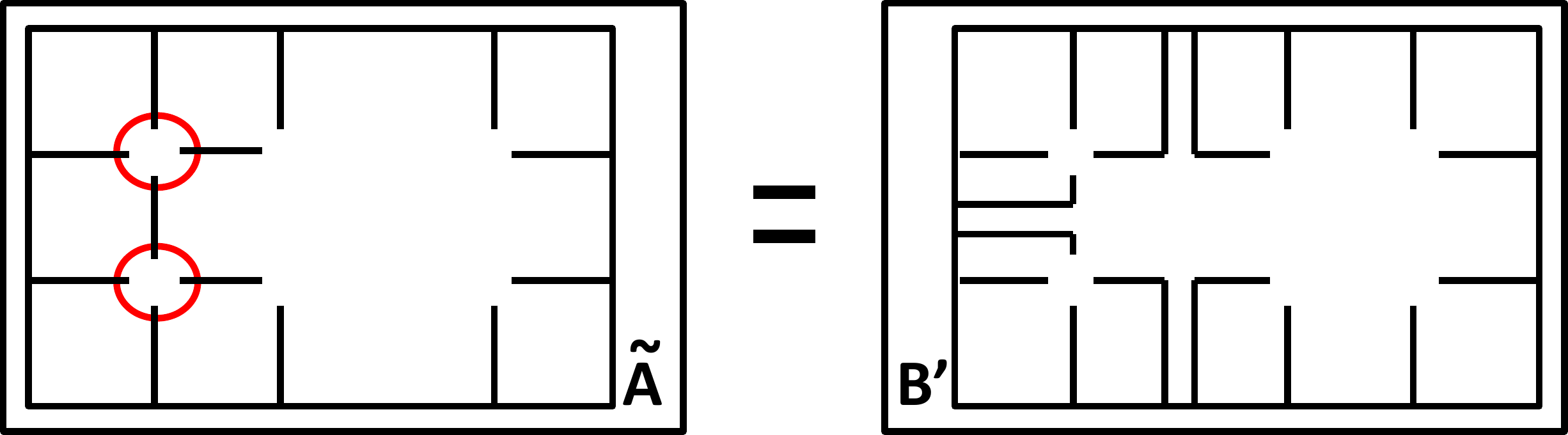}
}}\end{centering}
\end{equation}
where we have used the invariance of $B$ under a loop of MPO on the virtual boundary of the right plaquette.
Hence all states within the mutual ground space of neighboring plaquette terms are of the form
\begin{equation}
\begin{centering}\vcenter{\hbox{
\includegraphics[width=0.35\linewidth]{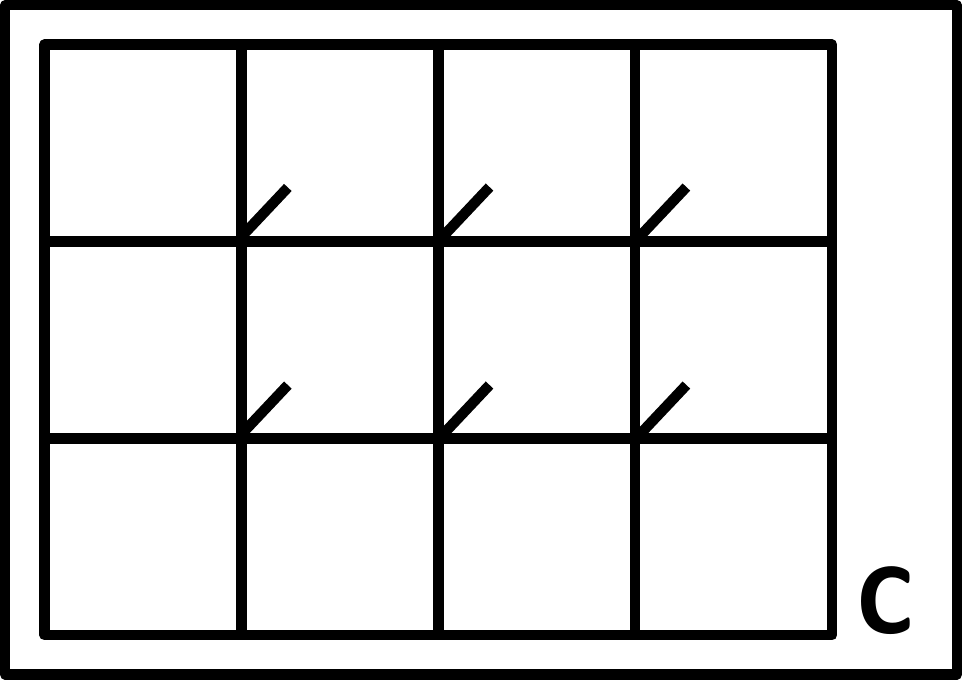}
}}\end{centering}
\end{equation}
for some boundary tensor $C$. 

It is possible to iterate this argument to show that the ground space of the parent Hamiltonian on any simply connected region with a nontrivial boundary is spanned by the PEPS on that region with an arbitrary virtual boundary tensor.

\textbf{Closure on a torus:} In the previous section we have shown that the ground state subspace on any simply connected region of the lattice with nontrivial boundary is spanned by a tensor network built from the PEPS tensor $\peps$ in the bulk and closed by an arbitrary tensor on the virtual boundary. 
If we proceed to close the region on a compact manifold, the additional plaquette terms of the parent Hamiltonian now crossing the boundary will further restrict the possible form of the boundary tensor in a way that depends on the topology of the manifold. 

We consider the specific case of closure on a $2\times 2$ torus below, and note that a direct generalization of the argument to any size of torus leads to the same conclusion.
By examining several different possible closures we refine the description of the boundary tensors that lead to a linearly independent set of ground states. We begin by looking at states in the intersection of two subspaces obtained from the following two different closures
\begin{equation}\label{c1}
\begin{centering}\vcenter{\hbox{
\includegraphics[width=0.8\linewidth]{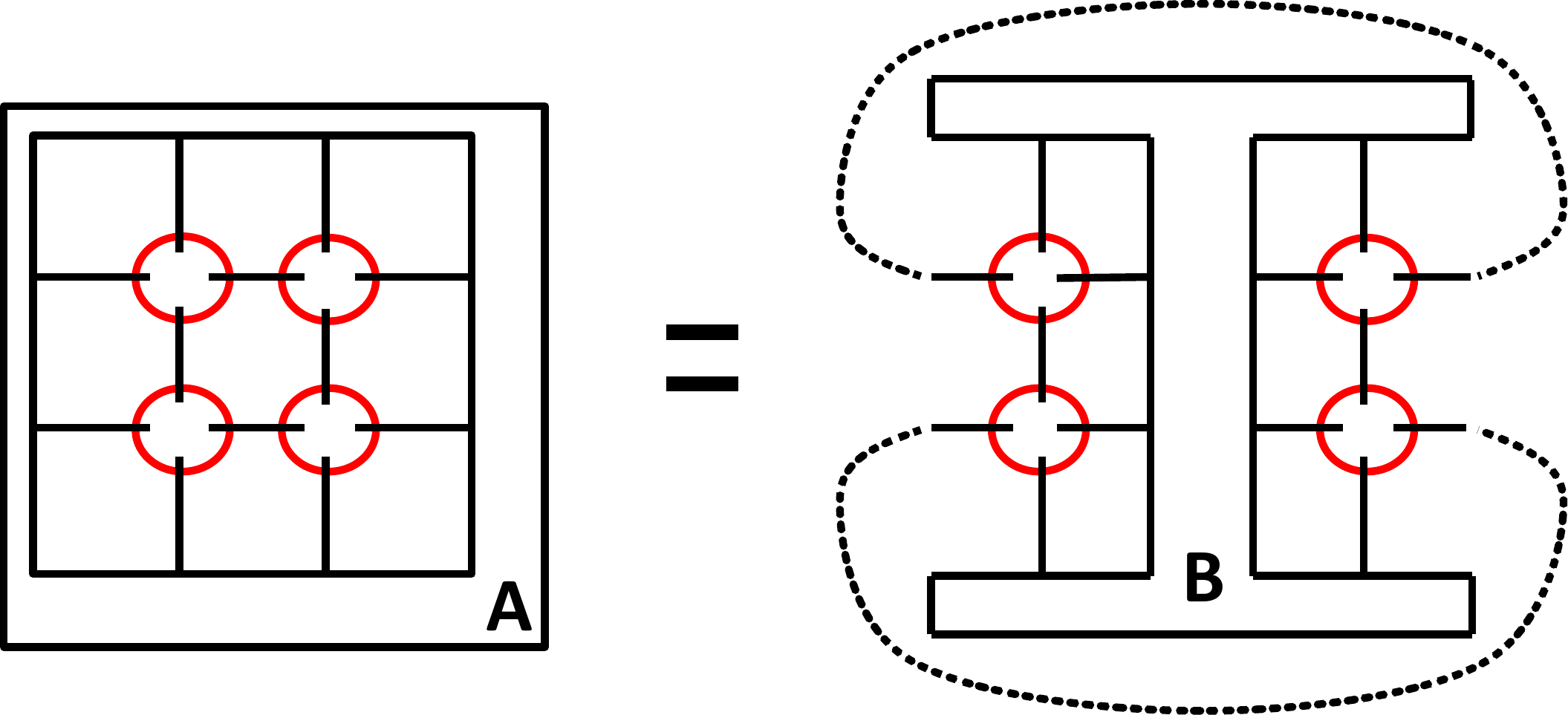}
}}\end{centering}
\end{equation}
which must be of this form, for some $A$ and $B$.
We utilize the pulling through condition twice and apply two generalized inverses to achieve
\begin{equation}
\begin{centering}\vcenter{\hbox{
\includegraphics[width=0.8\linewidth]{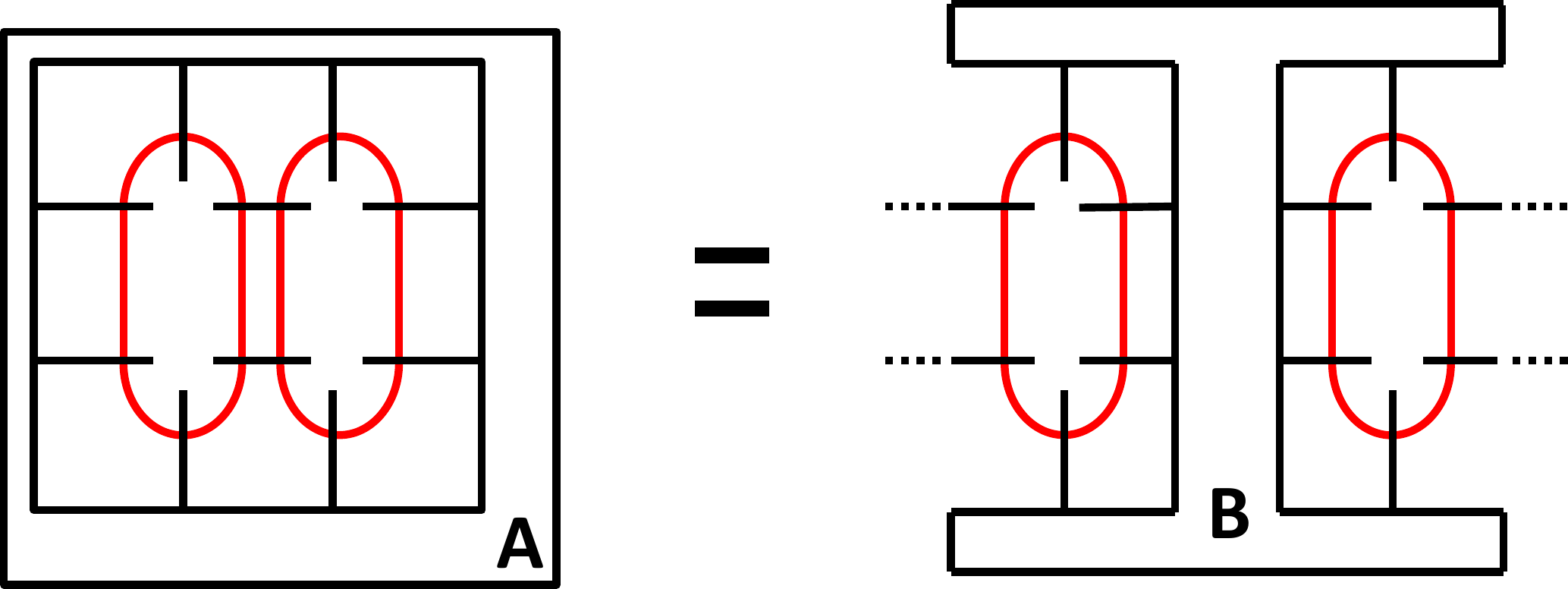}
}}\end{centering}.
\end{equation}
Next we apply a coarse-grained generalized inverse over two legs and find
\begin{equation}\label{c2}
\begin{centering}\vcenter{\hbox{
\includegraphics[width=0.8\linewidth]{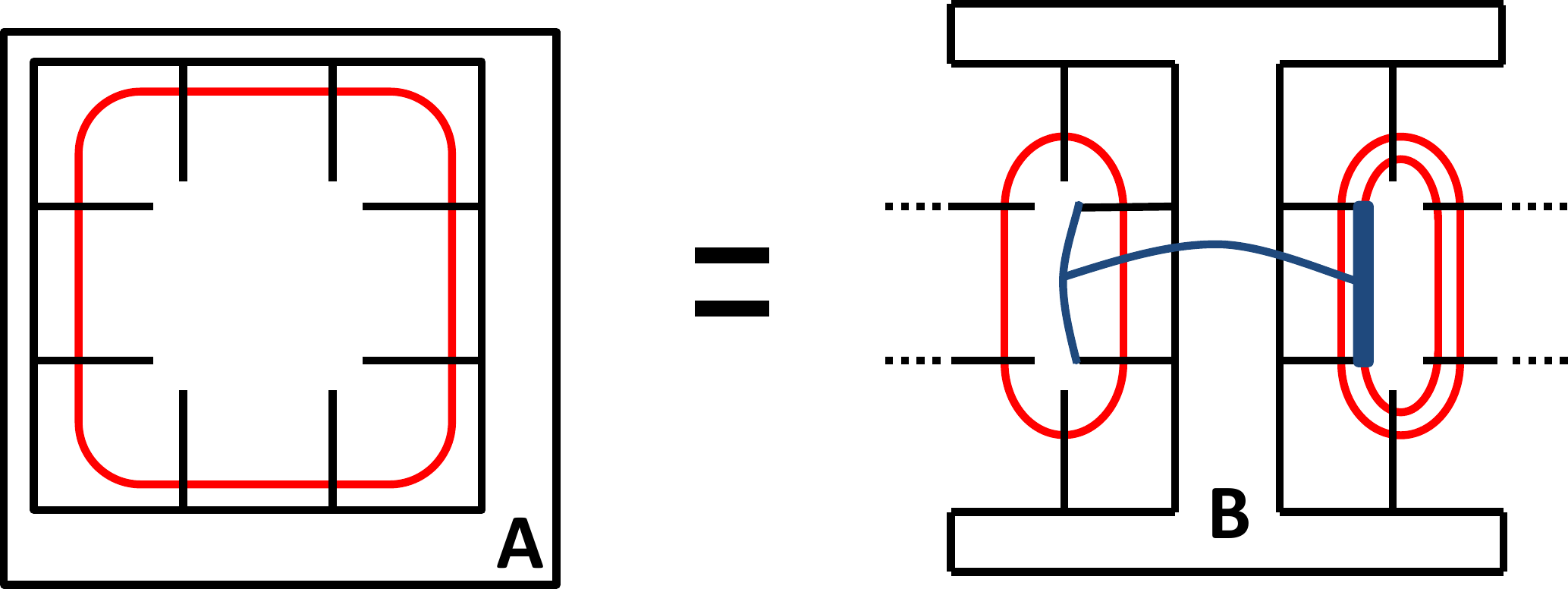}
}}\end{centering}.
\end{equation}
We note the following equality, attained after using coarse-grained pulling through twice,
\begin{equation}
\begin{centering}\vcenter{\hbox{
\includegraphics[width=0.8\linewidth]{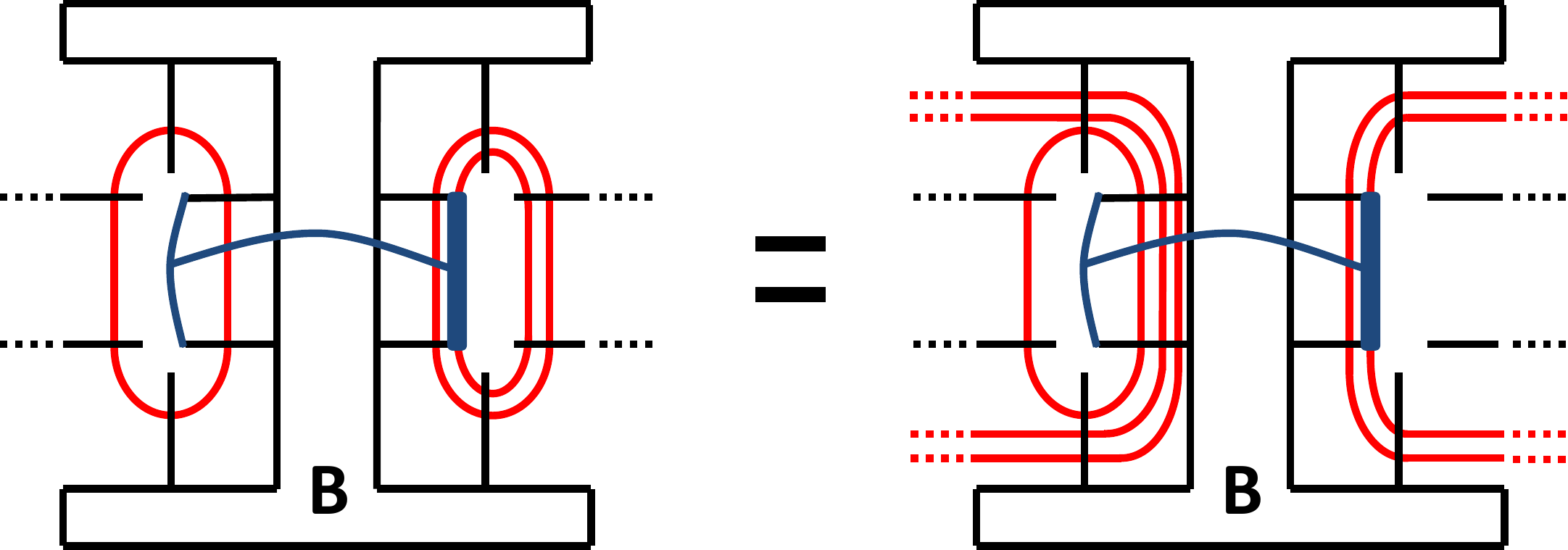}
}}\end{centering}
\end{equation}
and define the tensor $B'$,
\begin{equation}
\begin{centering}\vcenter{\hbox{
\includegraphics[width=0.8\linewidth]{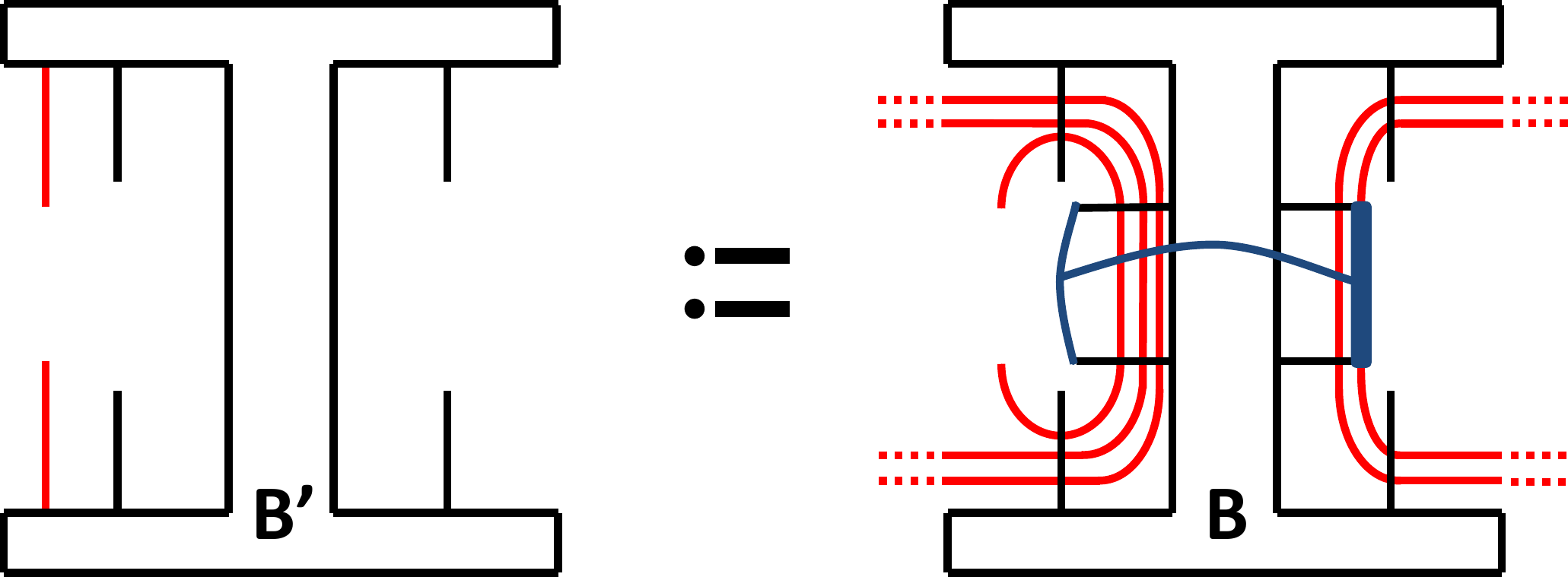}
}}\end{centering},
\end{equation}
for which we have the following equality
\begin{equation}\label{c3}
\begin{centering}\vcenter{\hbox{
\includegraphics[width=0.8\linewidth]{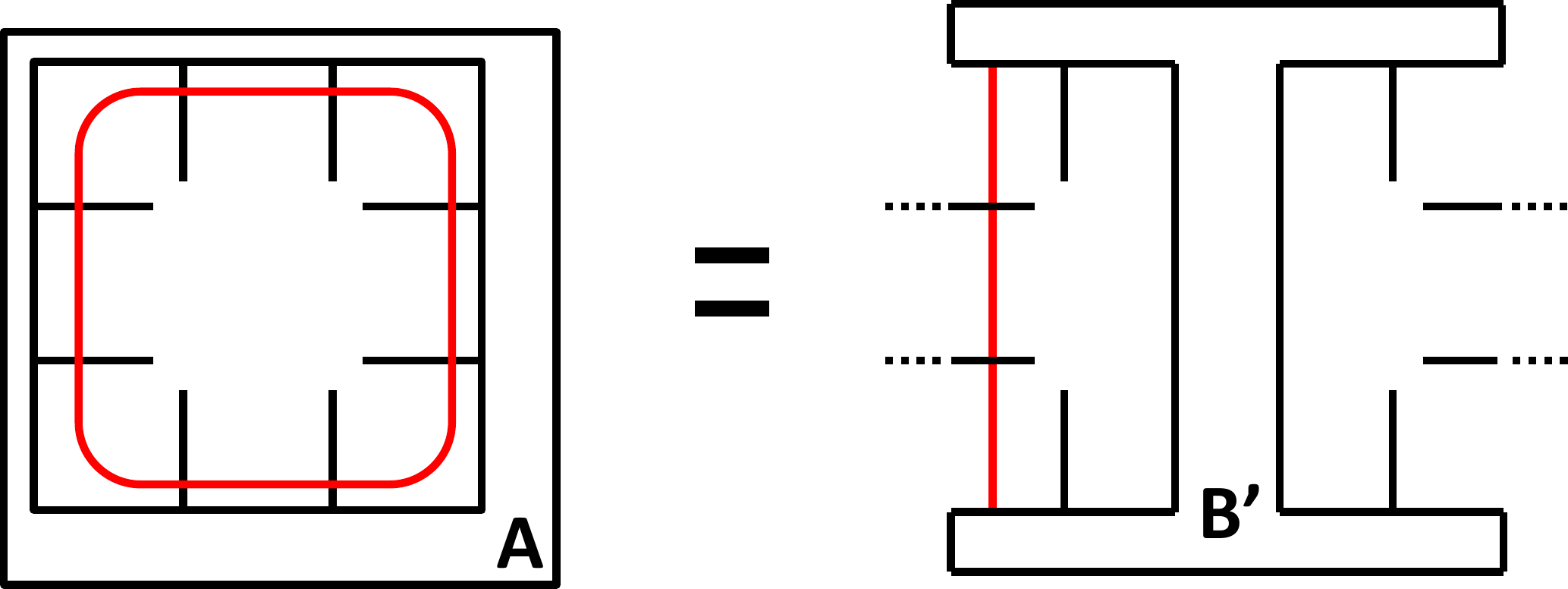}
}}\end{centering}.
\end{equation}

It is possible to repeat the preceding arguments to find states in the intersection of the $90^{\circ}$ rotated versions of the above boundary configurations. For states in the triple intersection we must have both Eq.~\eqref{c3} and the following
\begin{equation}\label{c4}
\begin{centering}\vcenter{\hbox{
\includegraphics[width=0.8\linewidth]{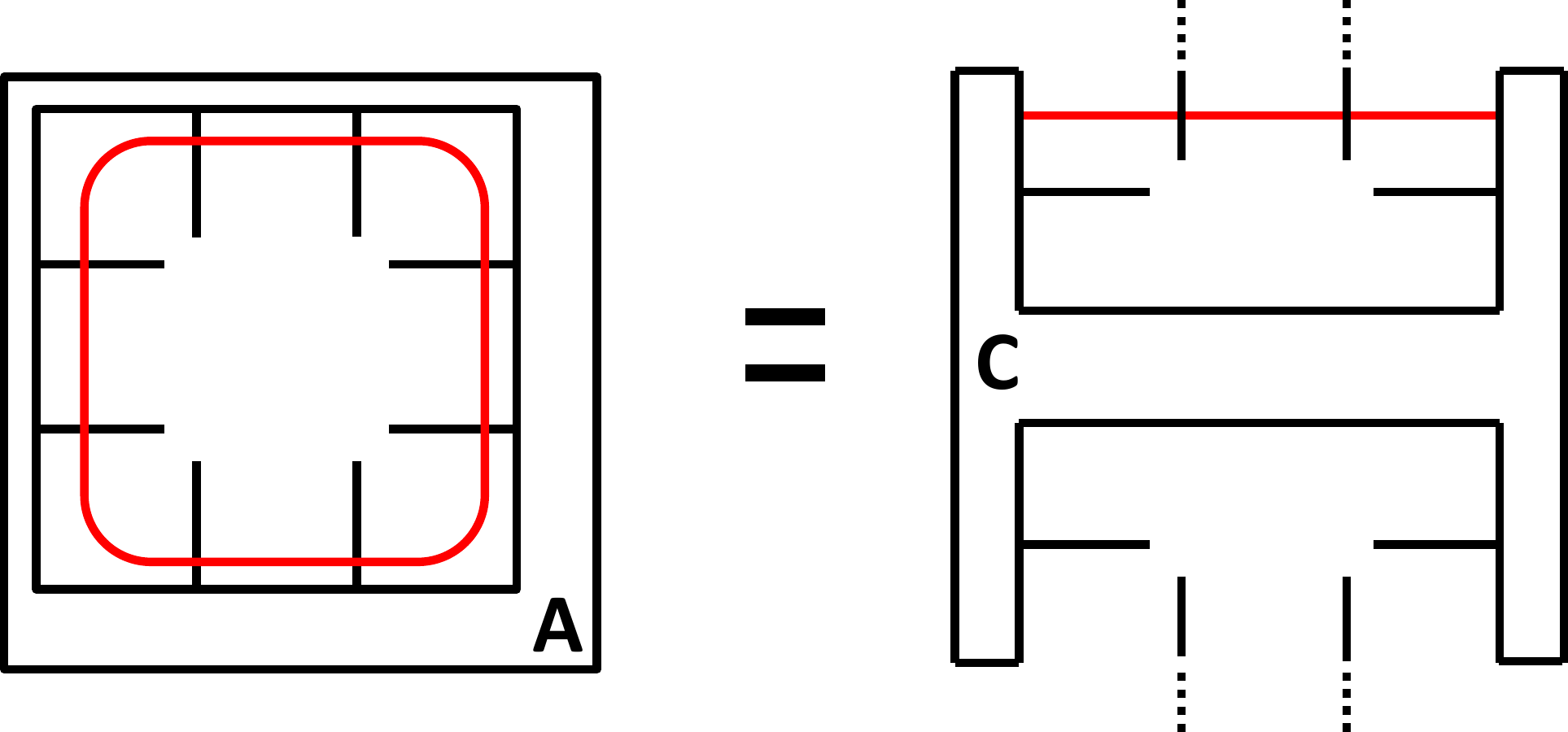}
}}\end{centering},
\end{equation}
and hence
\begin{equation}\label{c6}
\begin{centering}\vcenter{\hbox{
\includegraphics[width=0.8\linewidth]{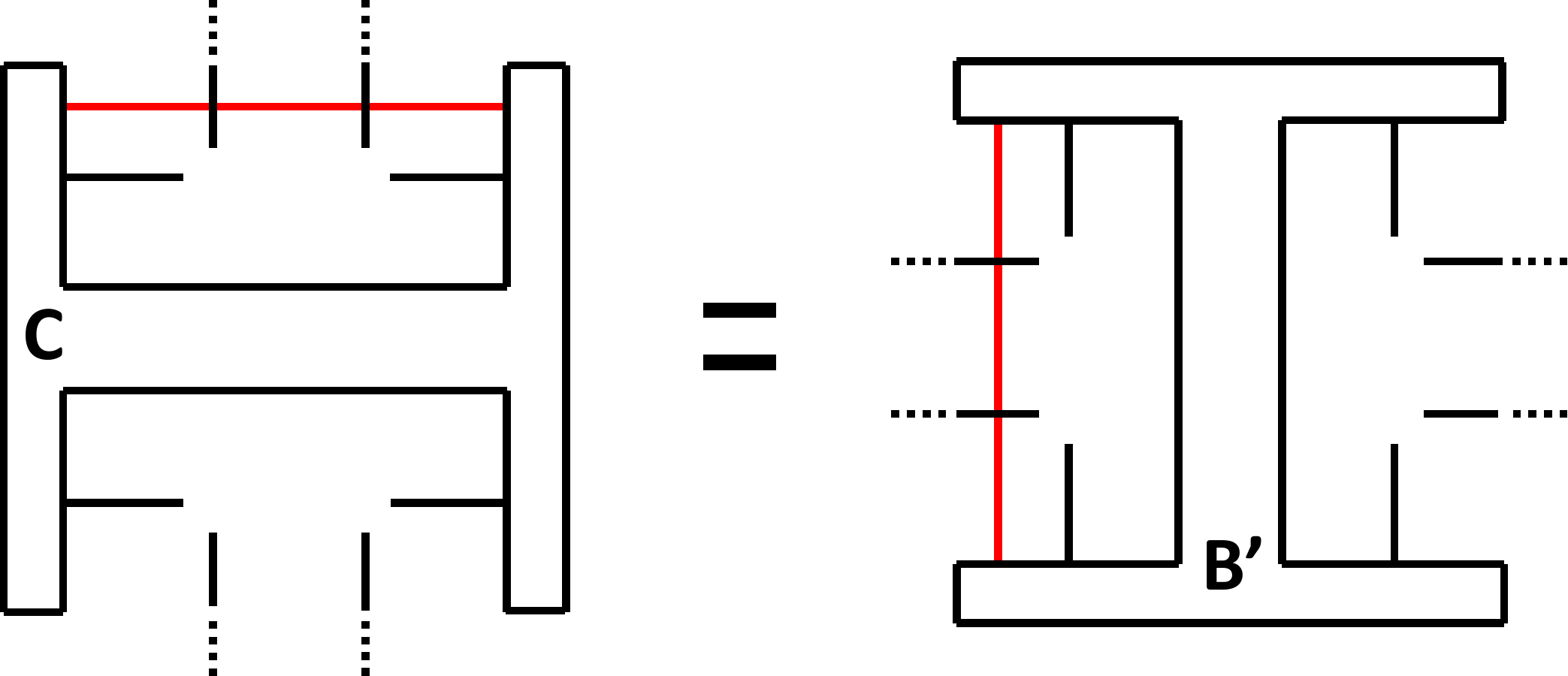}
}}\end{centering}
\end{equation}
for some $A$, $B'$ and $C$.
Viewing the tensors in Eq.~\eqref{c6} as linear maps from the vertical to horizontal indices we have
\begin{equation}
C_H C_V=B_H B_V
\end{equation}
where $C_V$ and $B_H$ are MPOs of length two. Writing $B_H^+$ for the pseudo-inverse of $B_H$ we have that 
\begin{equation}
C_H C_V=B_H \left( B_H^+ C_H \right) C_V
\end{equation}
and, defining $Q:= B_H^+ C_H$, the equality in Eq.~\eqref{c6} thus ensures the boundary tensor is of the following form
\begin{equation}\label{c7}
\begin{centering}\vcenter{\hbox{
\includegraphics[width=0.4\linewidth]{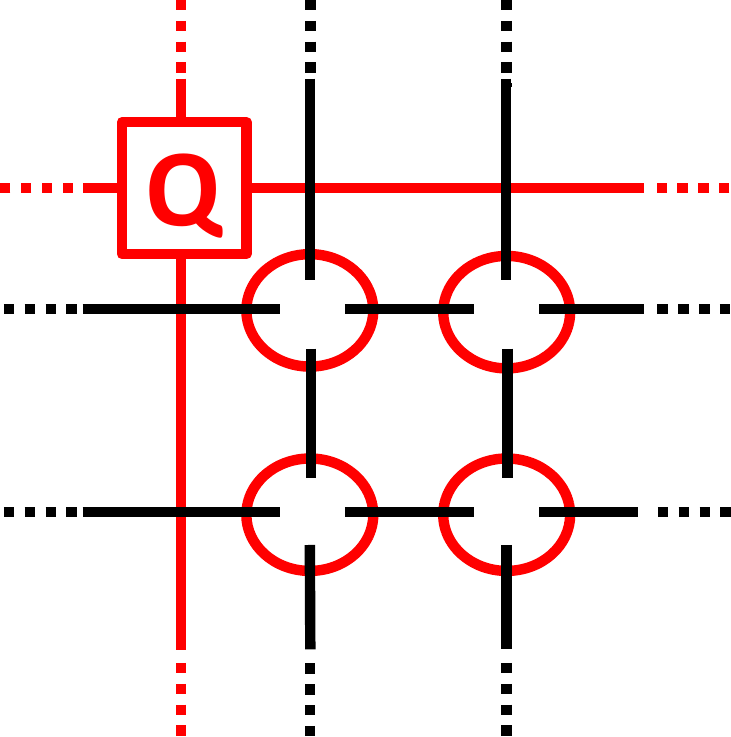}
}}\end{centering}.
\end{equation}
Repeating the above argument for the four different possible closures we have a set of equalities
\begin{equation}
\begin{centering}\vcenter{\hbox{
\includegraphics[width=0.8\linewidth]{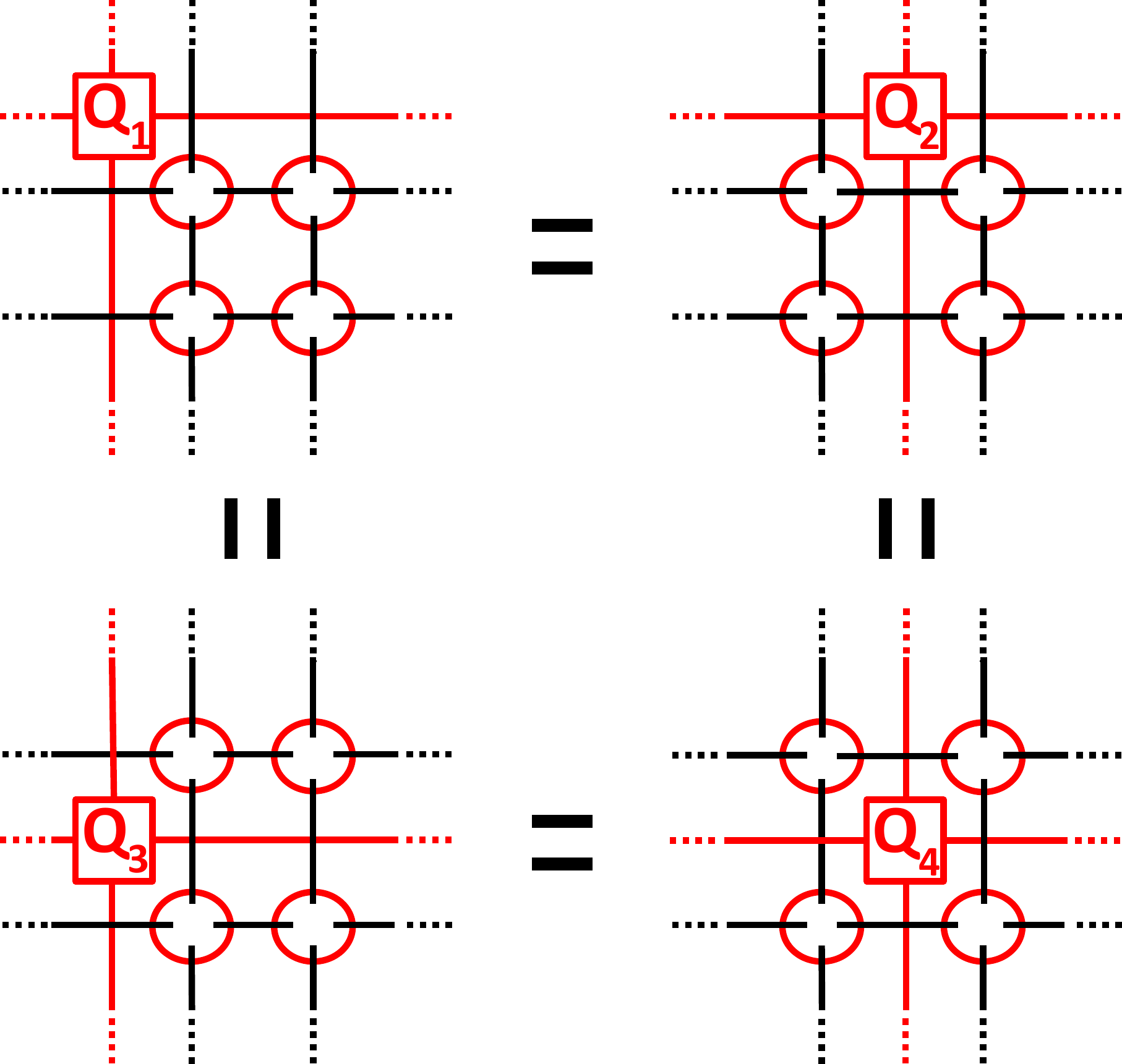}
}}\end{centering},
\end{equation}
for some possibly different boundary tensors $Q_i$.

\textbf{Ground state tensors:} The ground state tensors $Q$ are only defined up to transformations that do not affect the physical state. We first note that the equality of physical states for two different tensors $Q$ and $Q'$ on the same plaquette,
\begin{equation}\label{Qequiv1}
\begin{centering}\vcenter{\hbox{
\includegraphics[width=0.84\linewidth]{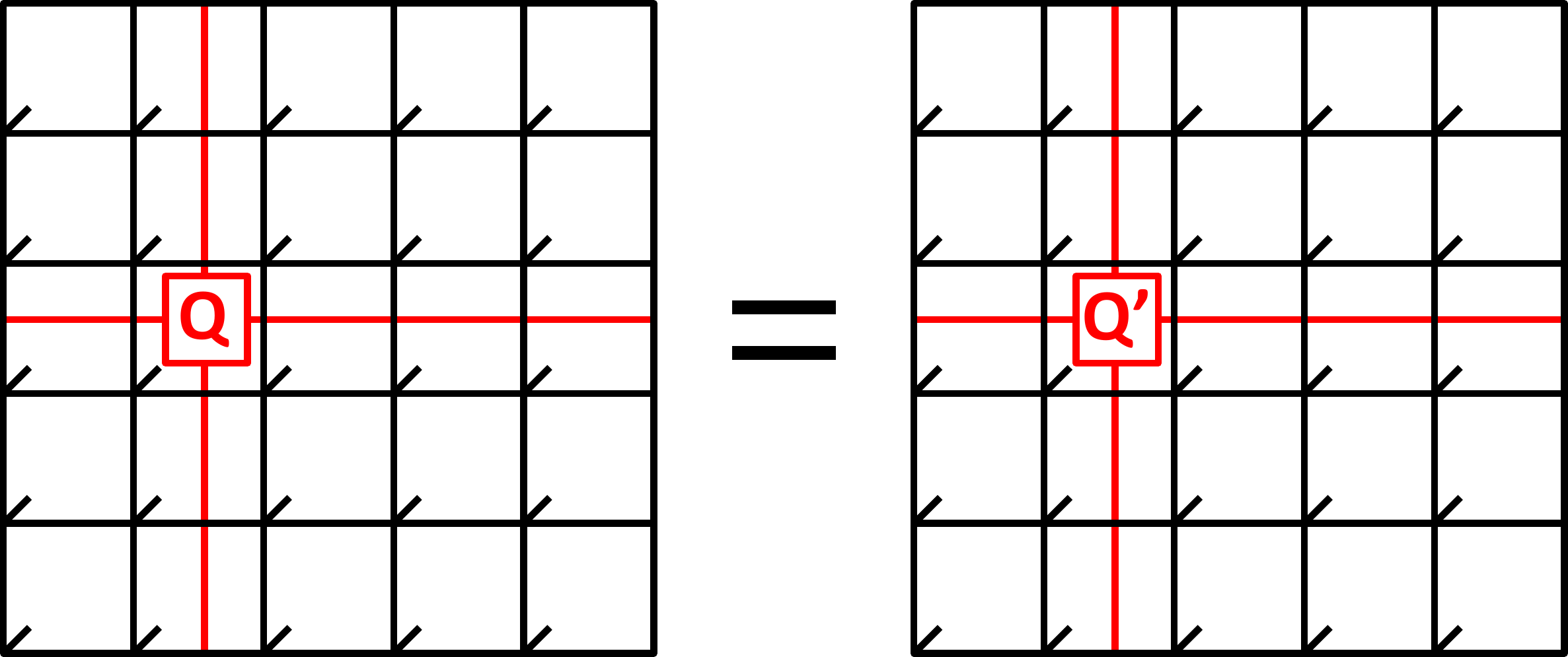}
}}\end{centering},
\end{equation}
is equivalent to equality of the physical states arising from the same tensors only involving the virtual bonds they directly act upon,
\begin{equation}\label{Qequiv2}
\begin{centering}\vcenter{\hbox{
\includegraphics[width=0.84\linewidth]{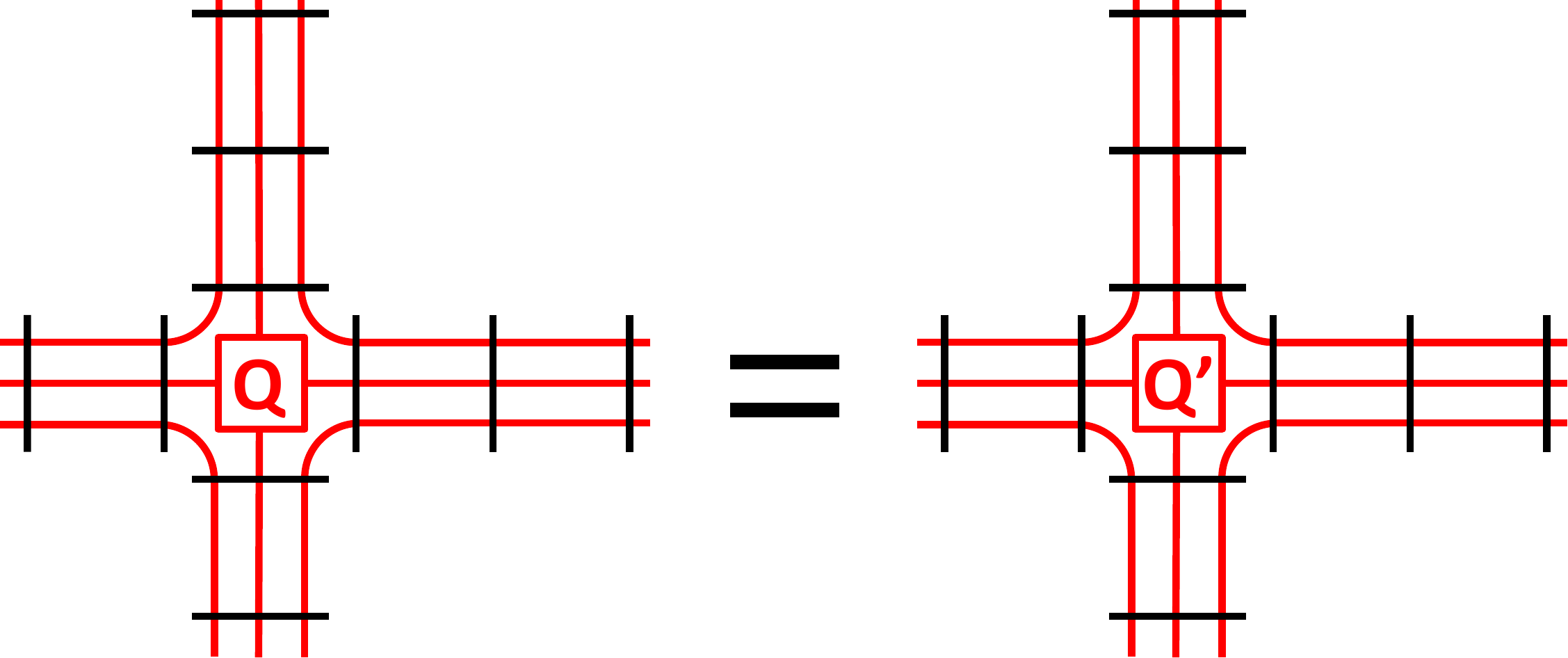}
}}\end{centering},
\end{equation}
by utilizing the pseudo-inverse on the topologically trivial region not acted upon by the MPO. We are assuming periodic boundary conditions in the above two equalities and for all lattices throughout the remainder of this section.

By further utilizing the RG moves we find that this is logically equivalent to equality of the states formed by $Q$ and $Q'$ on the smallest possible torus. Hence we use this condition [Eq.~\eqref{Qequiv3}] to define an equivalence relation on four index tensors, whose equivalence classes capture all tensors that lead to the same physical state,
\begin{equation}\label{Qequiv3}
\begin{centering}\vcenter{\hbox{
\includegraphics[width=0.84\linewidth]{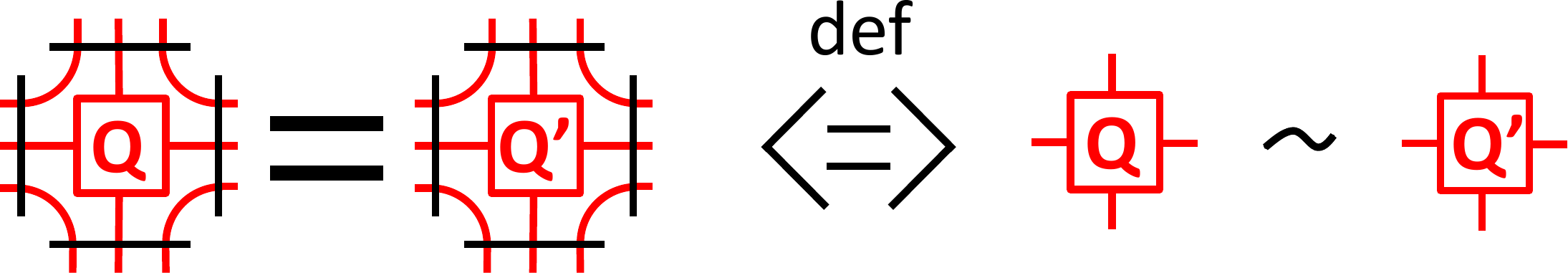}
}}\end{centering},
\end{equation}
note there are periodic boundary conditions for the left equality and open boundary conditions for the right equivalence relation.

By the arguments of the previous section we know that it is possible to close the PEPS tensor network, with a possibly site dependent $Q$ tensor, on any plaquette of the lattice to achieve the same physical state.
We now compare the closures at different points, first considering $Q$ tensors at two different locations along the same row of the dual lattice that give rise to the same physical state
\begin{equation}\label{Qequiv5}
\begin{centering}\vcenter{\hbox{
\includegraphics[width=0.85\linewidth]{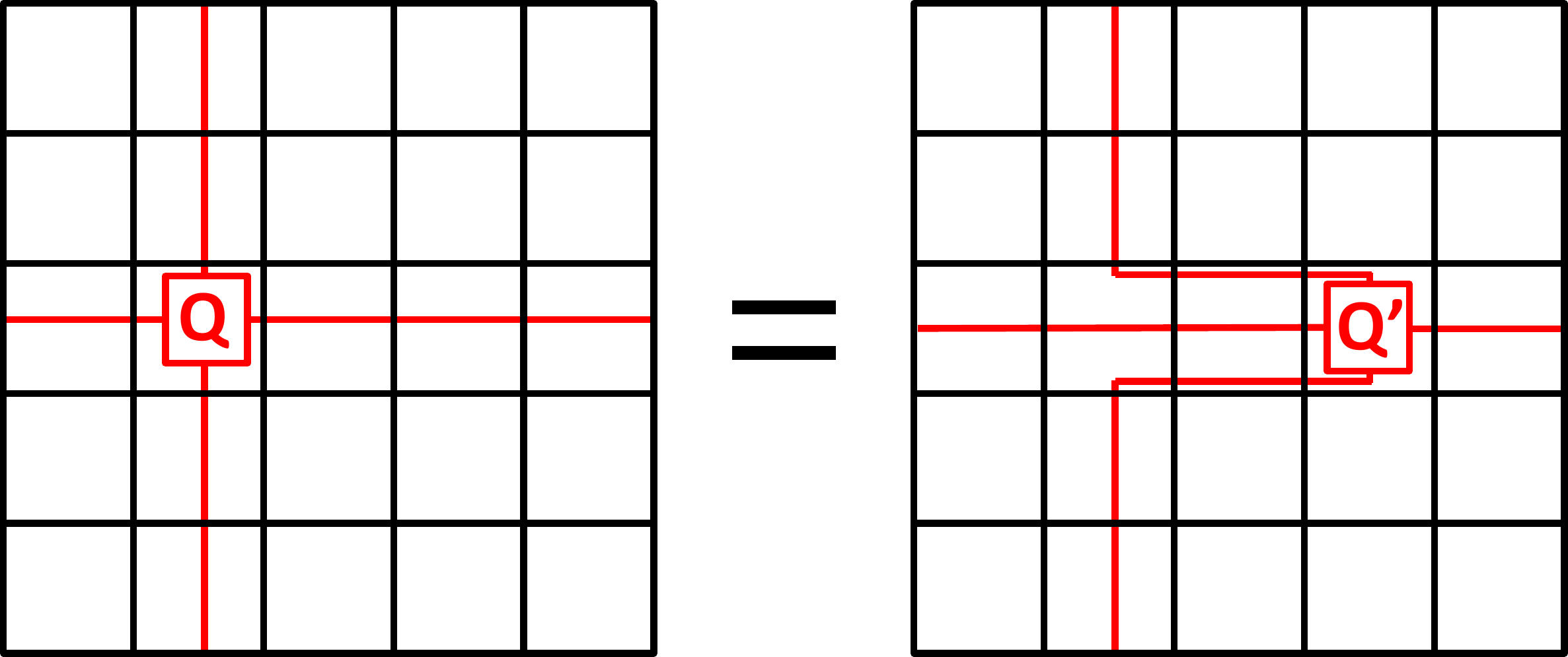}
}}\end{centering}
\end{equation}
where we have pulled through the MPO such that the boundary regions match. Now by employing the pseudoinverse on the bulk, followed by RG moves, we arrive at the equation
\begin{equation}\label{Qequiv6}
\begin{centering}\vcenter{\hbox{
\includegraphics[width=0.85\linewidth]{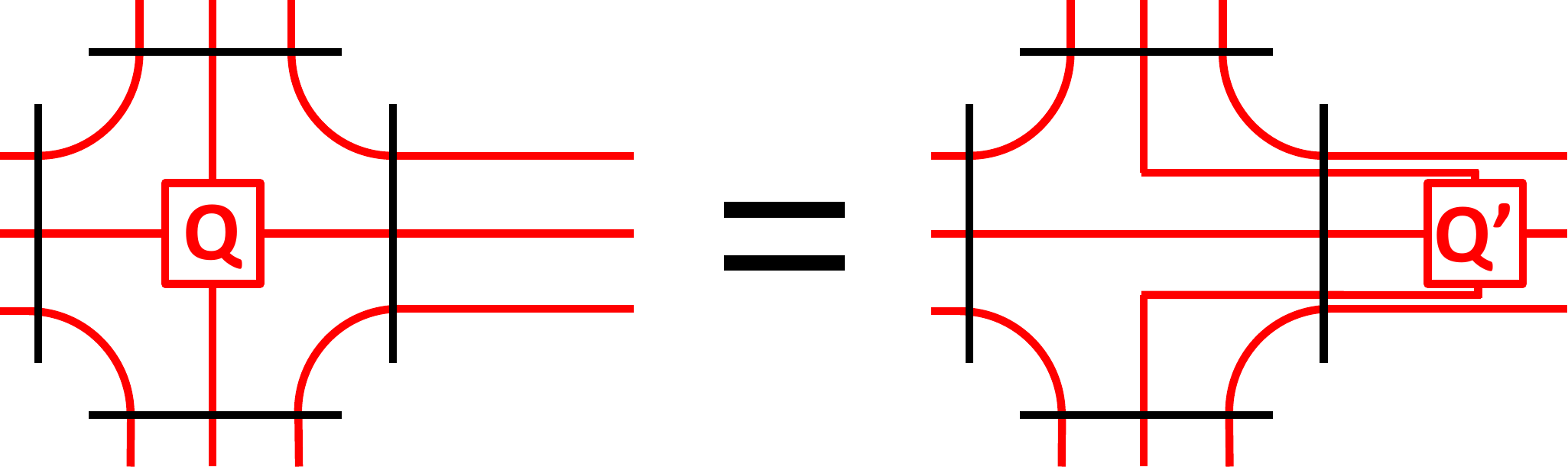}
}}\end{centering}
\end{equation}
which must be satisfied by $Q$ and $Q'$ if they give rise to the same physical state.

Now consider a third $Q''$ at a different plaquette along the same row, we proceed to compare each of the original tensors $Q,\ Q'$ to the new one $Q"$ via two different maps (constructed from RG moves) to arrive at two similar conditions
\begin{equation}\label{Qequiv7}
\begin{centering}\vcenter{\hbox{
\includegraphics[width=0.85\linewidth]{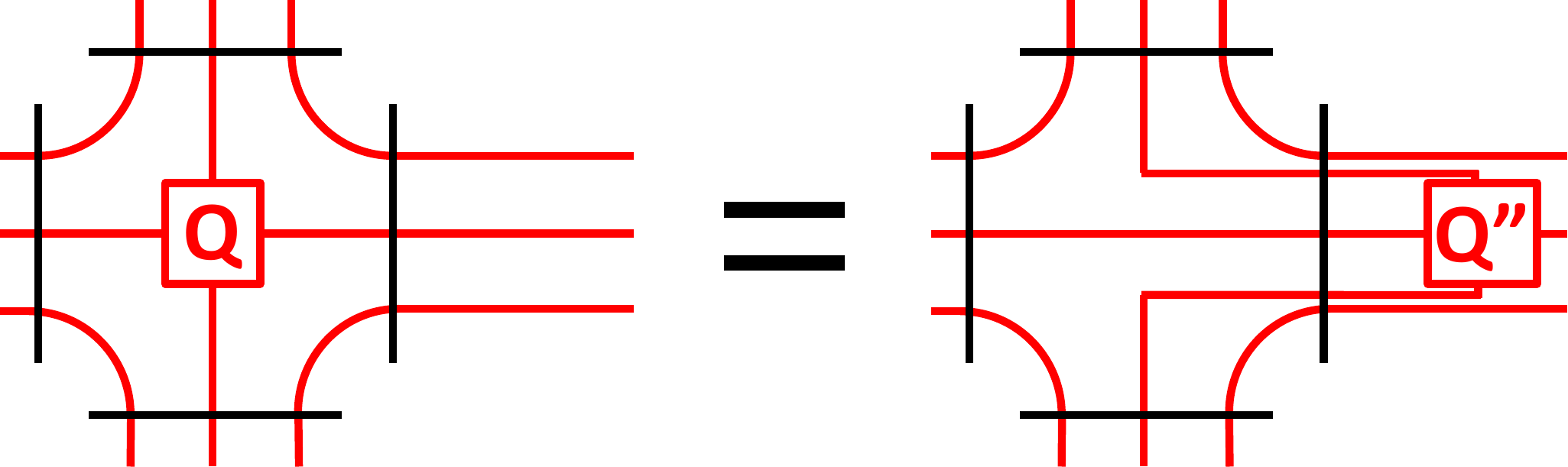}
}}\end{centering}
\end{equation}
and
\begin{equation}
\begin{centering}\vcenter{\hbox{
\includegraphics[width=0.85\linewidth]{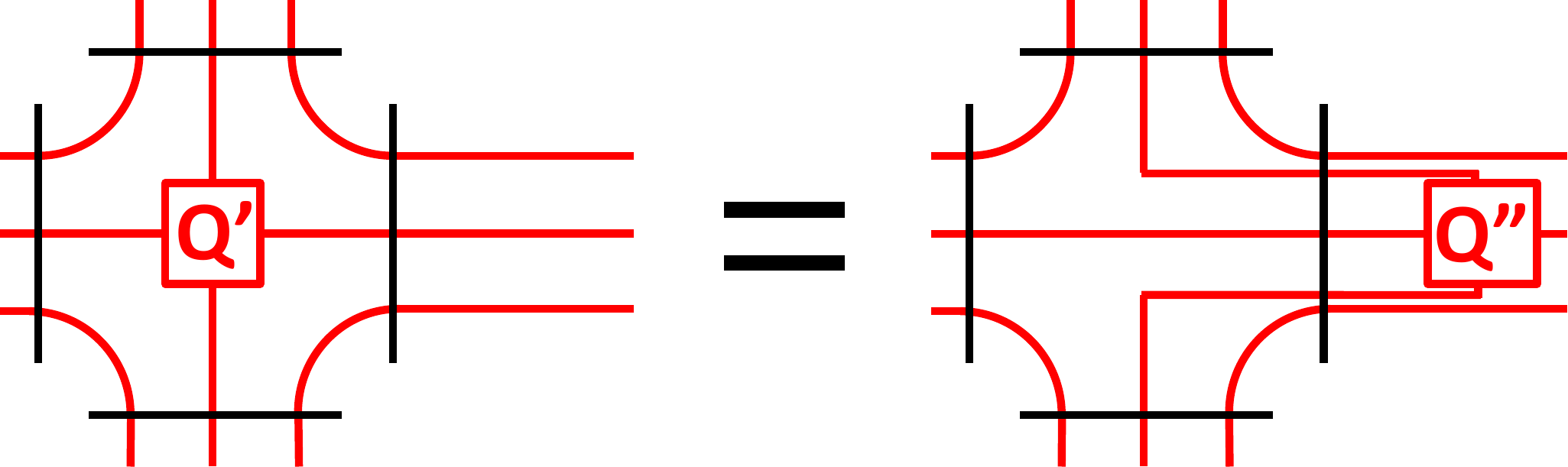}
}}\end{centering}
\end{equation}
which, together, imply equality of the two physical states that
arise from $Q$ and $Q'$ on the same plaquette of the PEPS, i.e., $Q\sim Q'$.
Note, a similar argument applies to boundary conditions shifted in the vertical direction, which then implies (in combination with the horizontal) that any two tensors closing the PEPS tensor network (possibly on different plaquettes) to give the same physical state must be equivalent. 

Hence on the level of equivalence classes we are searching for tensor solutions of the elementary pulling through equation [Eq.~\eqref{Ground state space}] in both the horizontal and vertical direction and it suffices to consider a particular representative $Q$  for each class. This ensures that the resulting tensor networks, with the same PEPS tensor on every site and a $Q$ tensor on the virtual level, will be translation invariant. To determine the degeneracy on the torus we must then look at the dimension of the subspace spanned by physical ground states coming from all the different $Q$ tensor solutions. Since the RG maps yield linear transformations between MPO injective PEPS on lattices of different sizes, which are invertible on the subspaces spanned by states of the form given in Eq.\eqref{Qequiv3} and Eq.\eqref{Qequiv1}, we can be sure that the exact degeneracy does not change for any finite system size. However, it is possible that as the system grows in size any number of 
states within the ground state subspace may converge to a single ground state or to zero in the thermodynamic limit. Hence one must examine the stability of the subspace as the system grows.

Finally we note that these closure arguments imply that any transformation preserving the ground state subspace can implicitly be rewritten as a transformation directly upon the $Q$ tensors, although there is no explicit formula in general.
 
\textbf{Tensor Network description of String-net Condensed States:} As first described in \cite{GuLevinSwingleWen09, 
BuerschaperAguadoVidal09}, the ground states of the string-net models have exact PEPS representations. 
Inside of every hexagon there is one virtual degree of freedom, these are connected to one another and 
to the degrees of freedom on the edges by tensors that sit on every vertex. The ground 
state is represented by the following tensor network
\begin{equation}\label{string-net on hexagon}
\begin{centering}\vcenter{\hbox{
\includegraphics[scale=0.3]{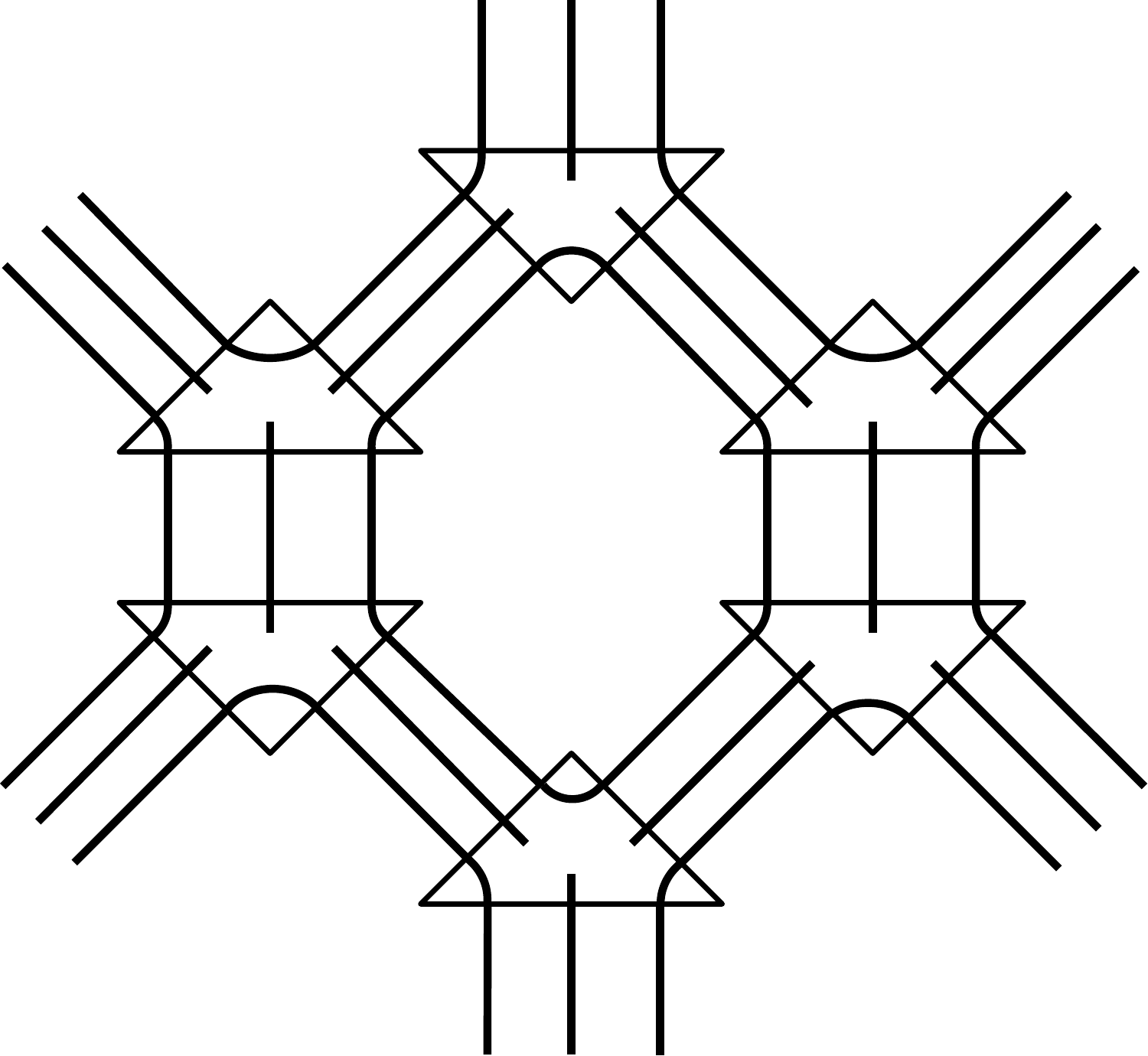}
}}\end{centering},
\end{equation}
where the tensor sitting on the vertices is 
\begin{equation}
\begin{centering}\vcenter{\hbox{
\includegraphics[width=0.6\linewidth]{8}
}}\end{centering},
\end{equation}
$d_s=v^2_s$ is the quantum dimension for sector $s$ and $D=\sqrt{\sum_s d^2_s}$ 
is the total quantum dimension. In the tensor network description, we make the convention that every 
closed loop comes with the multiplicative factor $a_s= d_s/D$ and the middle legs that connect each pair of tensors are copied to physical degrees of freedom on the adjacent vertices.
In the above expressions $G$ is a six index tensor, known as the symmetric $F$-symbol. For the sake of
completeness, we define these symbols and describe their symmetry properties 
which have been used in proving that the string-nets satisfy our axioms. The $F$-symbol is defined to be
a scalar map (when the branching is multiplicity free, i.e., 
$N_{ijk}$ is either $0$ or $1$) from one fusion path to 
another
\begin{equation}\label{F-symbol}
\begin{centering}\vcenter{\hbox{
\includegraphics[scale=0.2]{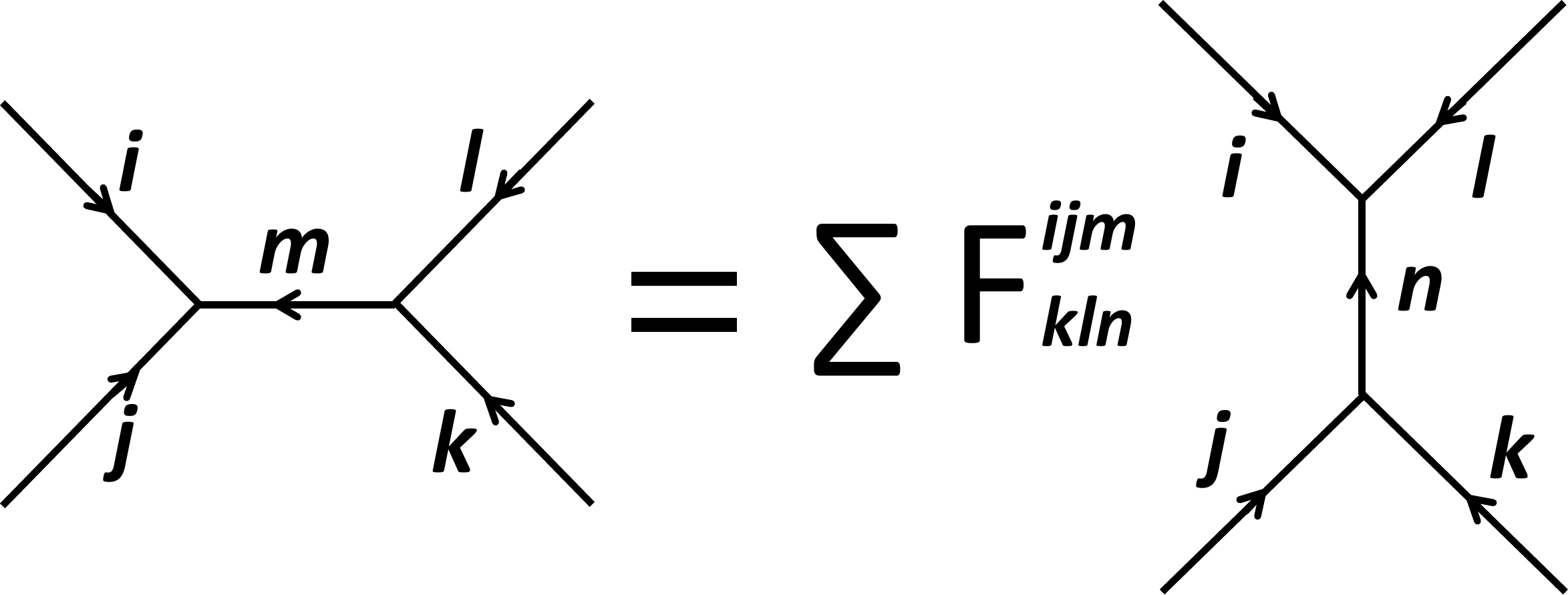}
}}\end{centering}\ .
\end{equation}
Then we define the $G$-symbol by the first line of the following expression
\begin{equation}\label{G in terms of F}
\begin{centering}\vcenter{\hbox{
\includegraphics[scale=0.25]{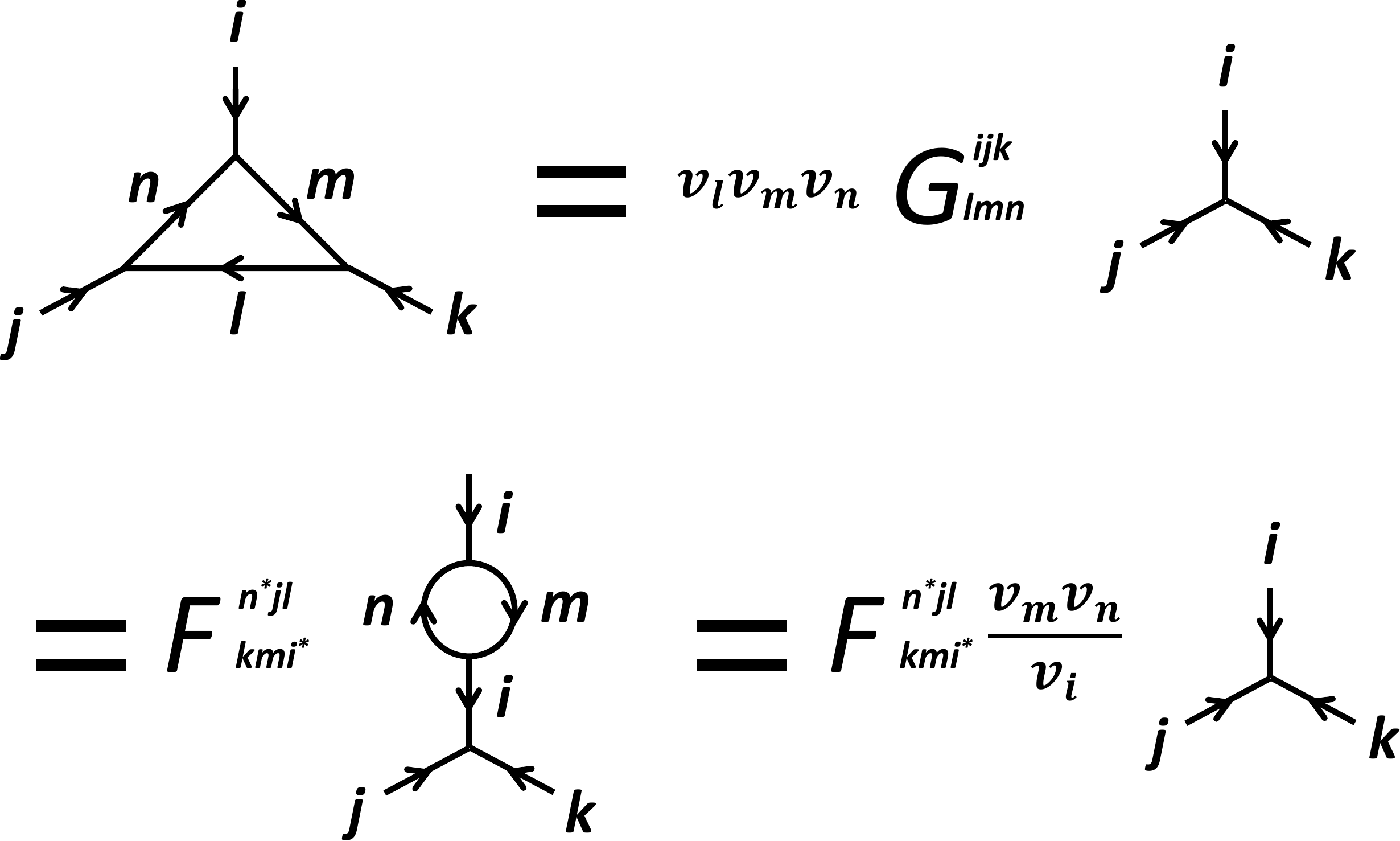}
}}\end{centering}\ .
\end{equation}
The second line of the above equation follows from the definition of the $F$-symbol and the fact that
\begin{equation}\label{line with buble}
\begin{centering}\vcenter{\hbox{
\includegraphics[width=0.5\linewidth]{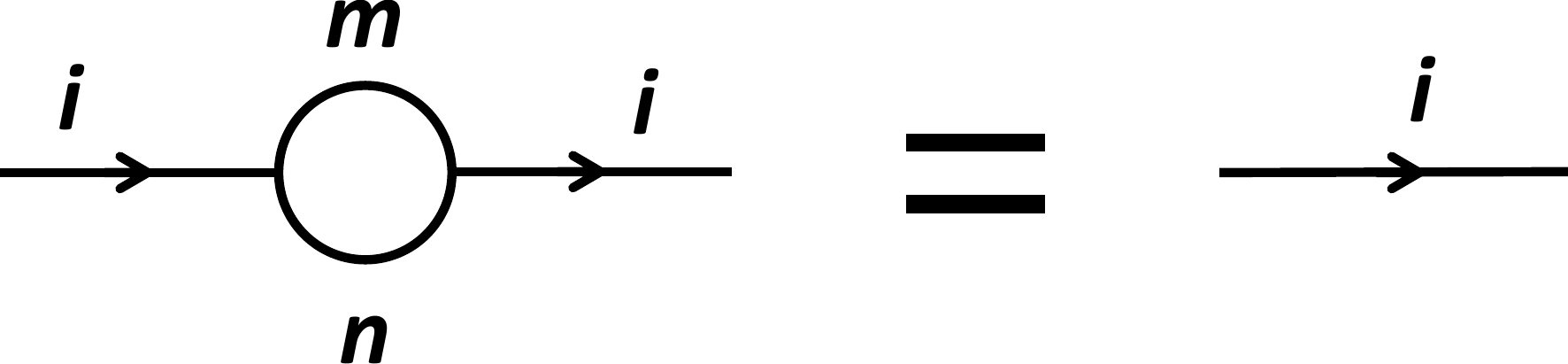}
}}\end{centering}
\end{equation}
regardless of arrow directions of $m$ and $n$. Graphically, the $G$-symbol corresponds to the following scalar map
\begin{equation}\label{G picture}
\begin{centering}\vcenter{\hbox{
\includegraphics[scale=0.2]{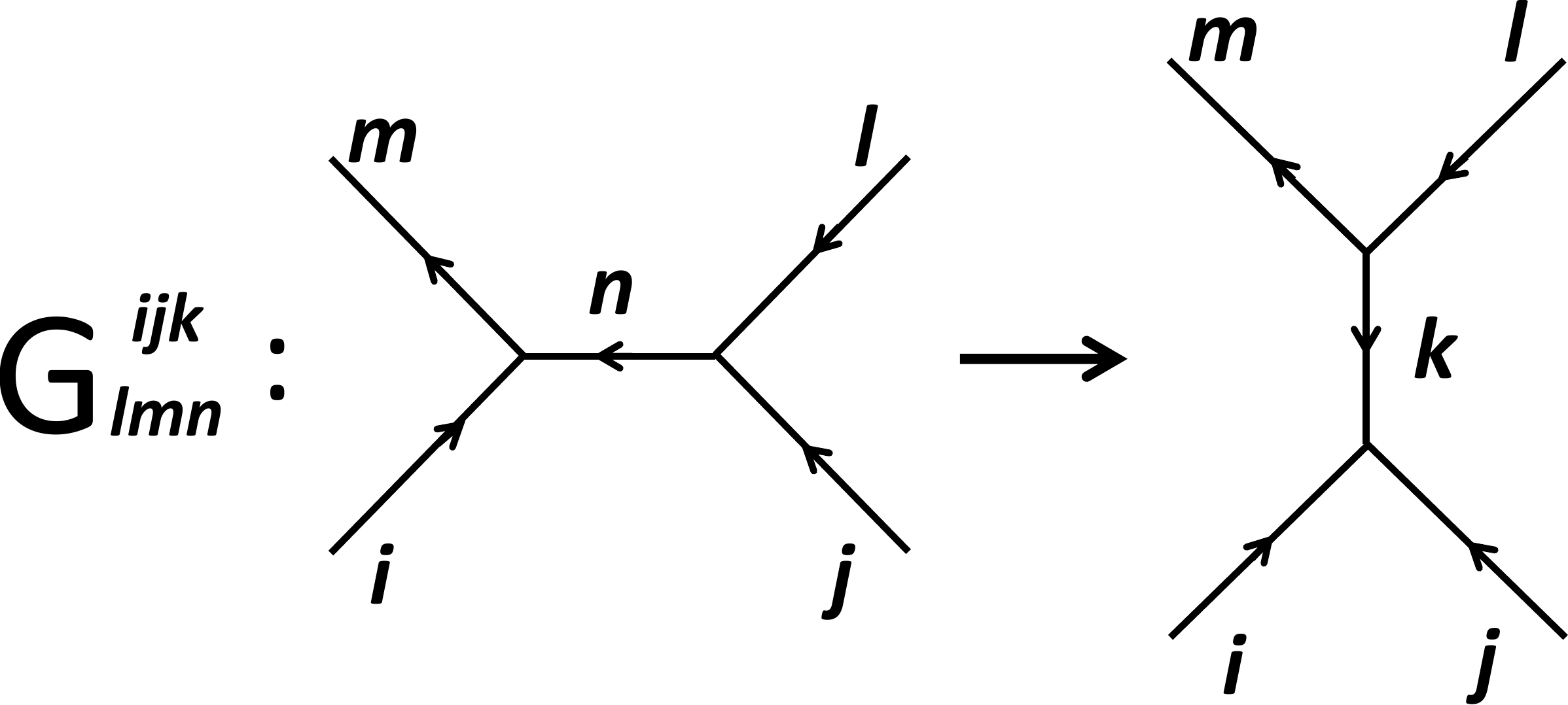}
}}\end{centering}\quad .
\end{equation}
By Eq.~\eqref{G in terms of F}, we find  $G^{ijk}_{\lambda \mu \nu}:= F^{\nu^* j \lambda}_{k \mu i^*}/ (v_\lambda v_i)$.
$F$ and $G$-symbols have nice symmetry properties, referred to as tetrahedral symmetry, 
shown in the following equations
\begin{equation}\label{Symmetry F}
F^{ijm}_{kln}= F^{lkm^*}_{jin}= F^{jim}_{lkn^*}= F^{imj}_{k^*nl} \dfrac{v_m v_n}{v_j 
v_l}.
\end{equation}
Using the above symmetry relations one finds that
\begin{equation}\label{gsymbol}
 G^{ijk}_{\lambda \mu \nu} = \frac{1}{v_k v_{\nu}} F^{i^*j^*k^*}_{\lambda^* \mu \nu^*}
\end{equation}
and clearly the $G$-symbols possess symmetry properties following from those in Eq.~\eqref{Symmetry F}.

The pentagon equation for these $G$-symbols follows from the pentagon equation for the $F$-symbols, and is given by
\begin{equation}\label{pentagoneqn}
 G^{i j k}_{\lambda \mu \nu}G^{i^* j^*  k^*}_{\alpha^* \beta^* \gamma^* } = \sum_n  d_n G^{k \alpha^* \beta}_{n \mu^* \lambda^*} G^{j \gamma^* \alpha}_{n\lambda^* \nu^*} G^{i \beta^* \gamma}_{n \nu^* \mu^*}\ .
\end{equation}

\textbf{Modular transformations on string-net PEPS:}  In this section we show that the modular transformations can be performed directly on the virtual
level of string-net PEPS on a torus. One can check that
the tensors $Q_{stu} = v_s v_tv_u G^{b^{*}du}_{t^{*}sa}G^{d^{*}bu^{*}}_{ts^{*}c}$, shown in Eq.~\eqref{Q-stringnet}, satisfy all the requirements on the ground state tensor of the string-net PEPS. 
\begin{equation}\label{Q-stringnet}
\begin{centering}\vcenter{\hbox{
\includegraphics[scale=0.2]{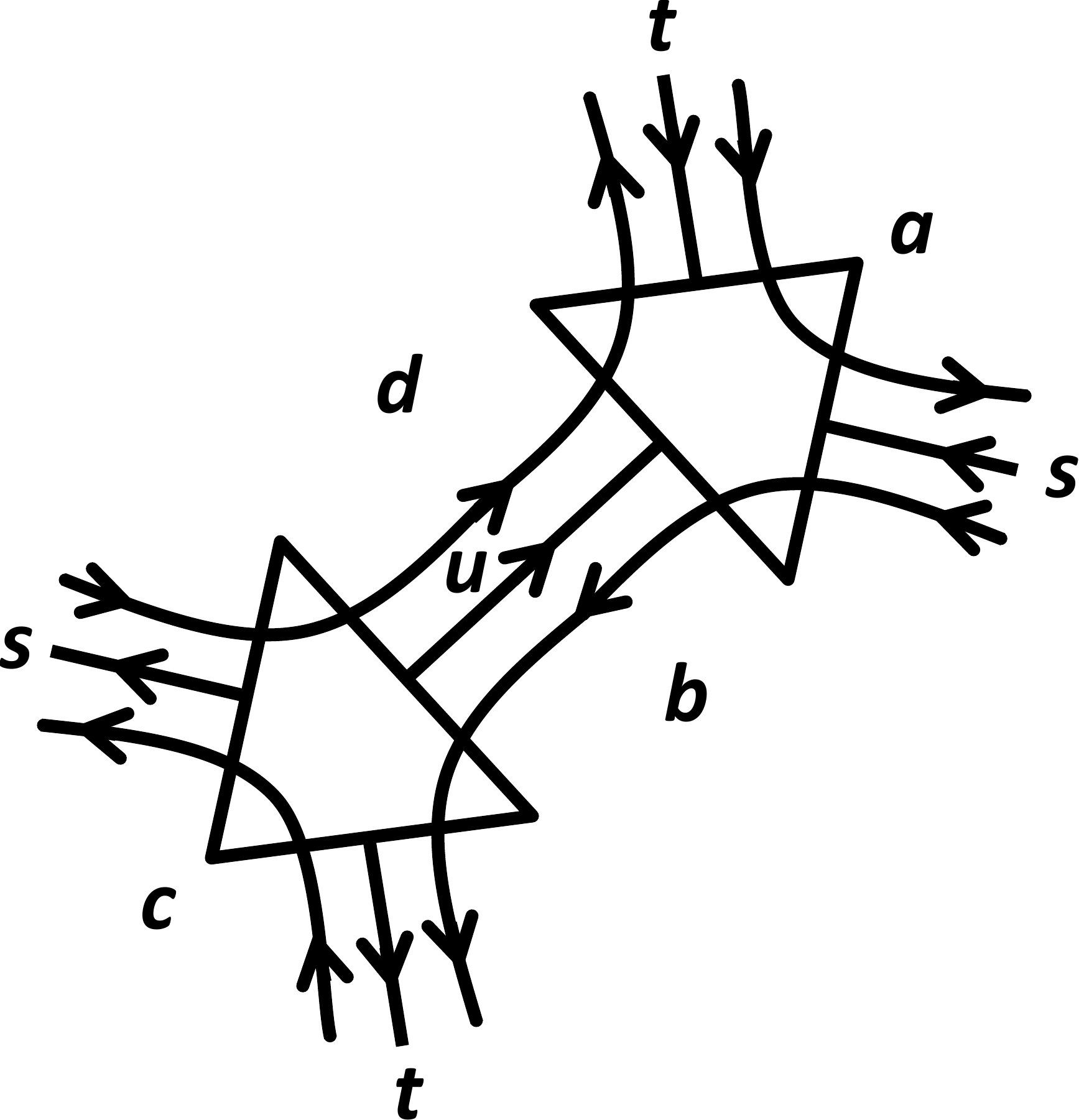}
}}
\end{centering}
\end{equation}
In Eq.~\eqref{Q-stringnet} the tensors are defined to be
\begin{equation}\label{MPOfuse}
\begin{centering}\vcenter{\hbox{
\includegraphics[width=0.6\linewidth]{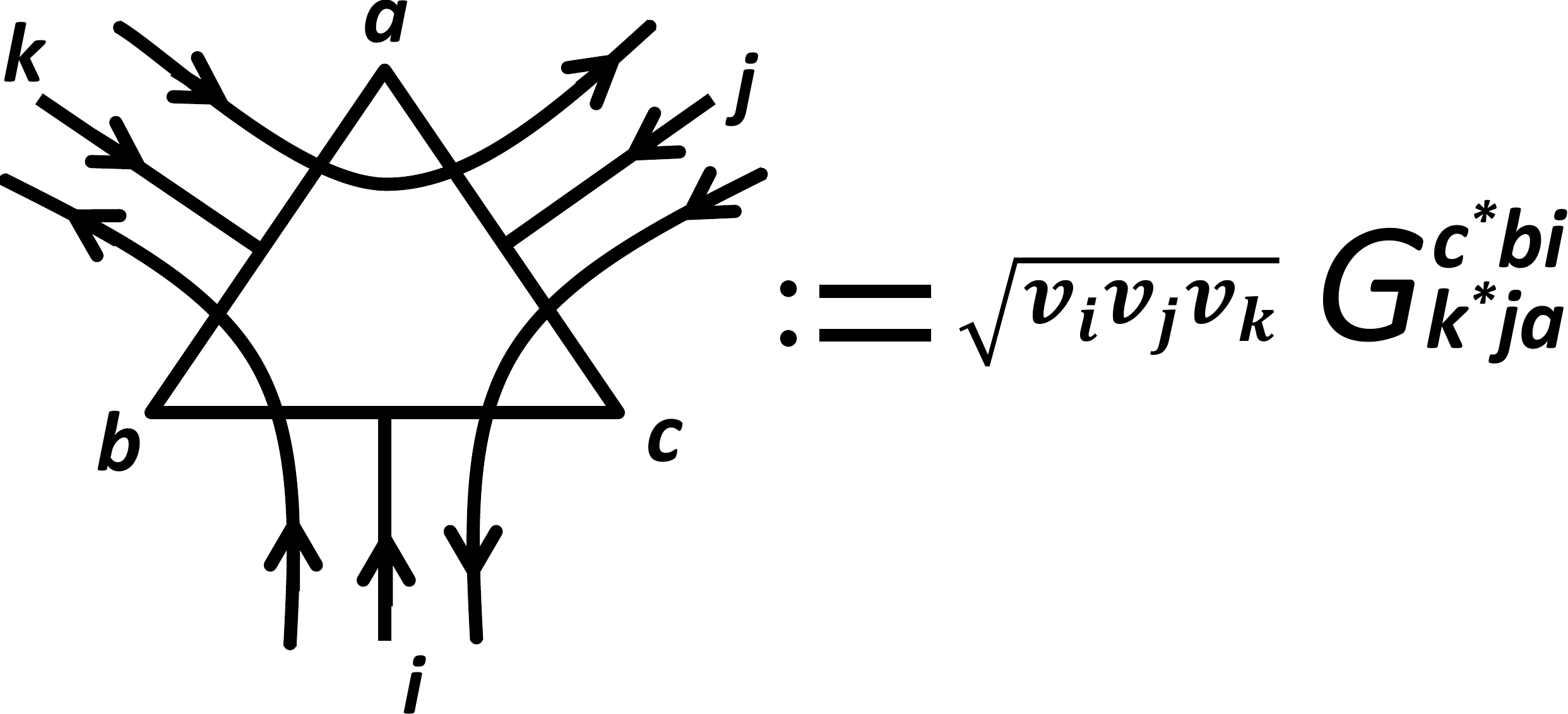}
}}\end{centering},
\end{equation}
the vertical indices $t$ must match since the PEPS is defined on a torus, and similarly for the horizontal indices $s$.

In general, $\left\{Q_{stu}\right\}$ will form an overcomplete set in the sense that not all $Q_{stu}$ will lead 
to linearly independent ground states. By utilizing the pentagon equation one can show that the $90^{\circ}$
rotated ground state tensor can be expressed as a linear combination of the original ground state 
tensors in the following way
\begin{equation}\label{P}
S(Q_{stu}) = \sum_n F^{stu}_{s^*t^*n} Q_{tsn}
\end{equation}
This agrees with the results of \cite{modular4}. 

By further utilizing the specific form of the $Q_{stu}$ tensors for the string-net PEPS [Eq.\eqref{Q-stringnet}] (and the pulling through condition [Eq.\eqref{string net - pulling through}]) 
one can express the ground state tensors of the ground states with a Dehn twist as a linear combination of the original 
ground state tensors via the explicit relation which we give here for completeness
\begin{equation}\label{P2}
Q^{\text{twisted}}_{stu} = \sum_n F^{stu}_{s^*t^*n}\ Q_{snt^*} \, ,
\end{equation}
where $t$ is the label wrapping around the torus in the direction of the twist in and $s$ is in the direction orthogonal 
to the twist, again agreeing with \cite{modular4}.

After determining the appropriate linear combinations 
of $Q_{stu}$ that lead to different ground states for a particular string-net model Eq.~\eqref{P} and Eq.~\eqref{P2} 
can be used to obtain the elements of the $S$ and $T$ matrices, respectively.

\end{document}